\documentclass[twocolumn]{aastex62}

\usepackage{graphicx}
\usepackage{amssymb}
\usepackage{xspace}
 \renewcommand{\thefootnote}{\fnsymbol{footnote}}

\begin{document}

\title{Identifying Tidal Disruption Events via Prior Photometric Selection of Their Preferred Hosts}

\author{K. Decker French}\altaffiliation{Hubble Fellow}% $^\dagger$}
\affil{The Observatories of the Carnegie Institution for Science, 813 Santa Barbara Street, Pasadena CA 91101}

\author{Ann I. Zabludoff}
\affil{Steward Observatory, University of Arizona, 933 North Cherry Avenue, Tucson AZ 85721}

%250 word limit
\begin{abstract}

A nuclear transient detected in a post-starburst galaxy or other quiescent galaxy with strong Balmer absorption is likely to be a Tidal Disruption Event (TDE). Identifying such galaxies within the planned survey footprint of the Large Synoptic Survey Telescope (LSST)---before a transient is detected---will make TDE classification immediate and follow-up more efficient. Unfortunately, spectra for identifying most such galaxies are unavailable, and simple photometric selection is ineffective; cutting on ``green valley'' UV/optical/IR colors produces samples that are highly contaminated and incomplete. Here we propose a new strategy using only photometric optical/UV/IR data from large surveys. Applying a machine learning Random Forest classifier to a sample of $\sim$400k SDSS galaxies with {\it GALEX} and {\it WISE} photometry, including 13,592 quiescent Balmer-strong galaxies, we achieve 53--61\% purity and 8--21\% completeness, given the range in redshift. For the subset of 1299 post-starburst galaxies, we achieve 63--73\% purity and 5-12\% completeness. Given these results, the range of likely TDE and supernova rates, and that 36--75\% of TDEs occur in quiescent Balmer-strong hosts, we estimate that 13--99\% of transients observed in photometrically-selected host galaxies will be TDEs and that we will discover 119--248 TDEs per year with LSST. Using our technique, we present a new catalog of 67,484 candidate galaxies expected to have a high TDE rate, drawn from the SDSS, Pan-STARRS, DES, and {\it WISE} photometric surveys. This sample is 3.5$\times$ larger than the current SDSS sample of similar galaxies, thereby providing a new path forward for transient science and galaxy evolution studies.

\end{abstract}

\section{Introduction}

%\footnotetext[0]{$^\dagger$ Hubble Fellow}
\renewcommand*{\thefootnote}{\arabic{footnote}}

The bimodality between blue starforming galaxies and red passive galaxies has been long studied, with galaxies expected to transition to the so-called ``red-sequence'' after they stop forming stars. Post-starburst galaxies, characterized by their strong Balmer absorption and weak emission lines,  are thought to mark a transition between these two types of galaxies \citep{Dressler1983, Couch1987, Zabludoff1996, Yang2004, Yang2008, Wong2012} and have intermediate ``green'' colors. However, the ``green valley'' of galaxies is also populated by slowly evolving galaxies, AGN, dusty star-forming galaxies, and rejuvenated red galaxies. As such, although most  post-starburst galaxies lie in the optical green valley \citep{Wong2012}, spectroscopy has still been required to select them without significant contamination.

The ability to select post-starburst galaxies from next generation surveys such as LSST \citep{Ivezic2008} and ZTF \citep{Bellm2014} is important not only for galaxy evolution studies, but also for transient science, particularly for the discovery of tidal disruption events (TDEs).  Post-starburst and other quiescent Balmer-strong galaxies host tidal disruption events \citep{Arcavi2014, French2016, French2017,Graur2017, Law-Smith2017} at rates 20-200$\times$ higher than expected given the rarity of such galaxies. While over $\sim10$ TDEs are expected to be discoverable each night with LSST \citep{vanvelzen2011}, they must be found among the ten million other transient event detections that night. The preference for unusual host galaxies provides a way to know which transients are likely to be TDEs, even before the transient is detected.

Early time lightcurves of TDEs are essential for fitting the black hole properties of the event, especially the black hole masses \citep[e.g.,][]{Guillochon2017, Mockler2018}. Yet, identifying likely TDEs is complicated; even among the small number of TDEs classified to date the color evolution \citep{Holoien2016}, time evolution \citep{Blagorodnova2017}, and peak magnitude \citep{Blagorodnova2017} vary. Deviations among their lightcurves are expected from theory as well, e.g., for a TDE around a black hole binary \citep{Coughlin2016}. The many TDEs found by LSST will exhibit an even wider variety of observational signatures. Using the host galaxy preference as a tool for identifying TDEs in LSST and other transient surveys will hasten their discovery and allow for rapid follow-up.

How then do we identify such potential host galaxies? Many of the events found by LSST will be in the southern hemisphere, beyond the reach of galaxy spectroscopic surveys such as the Sloan Digital Sky Survey (SDSS, \citealt{Strauss2002}). Therefore, we must develop a new approach to find quiescent Balmer-strong galaxies, as well as the subset that are post-starburst galaxies and even more preferred by TDEs \citep{Arcavi2014,French2016}.  Here, we present a machine learning method for the photometric selection of post-starburst and quiescent Balmer-strong galaxies based solely on their ultraviolet, optical, and infrared colors and color gradients, i.e., data of the kind that will be available for the host galaxies of transients discovered with LSST. We characterize the performance of this method using metrics of purity (the fraction of all photometrically selected galaxies that are spectroscopically confirmed) and completeness (the fraction of all spectroscopically confirmed galaxies that are photometrically selected ). In \S2, we describe the datasets used for training and testing the machine learning classifier as well as our methodology. In \S3, we characterize the performance of our photometric-only selection, consider the implications for transient follow-up strategies, and present a large new catalog of photometrically-identified galaxies. In \S4, we discuss our assumptions and the impact of future surveys. In \S5, we summarize our conclusions. When needed, we assume a cosmology of $\Omega_m=0.3$, $\Omega_\Lambda=0.7$, and $h=0.7$.

\section{Method}

\subsection{Training Set Data}
\label{sec:data}

We select a parent galaxy sample from the SDSS DR12 main galaxy survey \citep{Strauss2002, alam2015} to use for training and testing our selection method. We require the spectra to have median SNR$>10$, and the resulting sample ranges from $10-98$. We select for good H$\alpha$ equivalent width measurements by requiring {\tt h\_alpha\_eqw\_err > 1}. We restrict our sample to galaxies with $z>0.01$ in order to limit the effects of aperture bias when galaxies are much larger than the 3\arcsec\ SDSS fibers.

We use optical $ugriz$ photometry from SDSS, using the {\tt modelmag} magnitude, which is known to provide stable colors while containing most of the galaxy light \citep{Al.2004}. $NUV$ magnitudes are selected for galaxies matching the SDSS galaxies within 4\arcsec\ from the {\it GALEX} GCAT catalogs. This radius is similar to the FWHM of the $NUV$ PSFs and much larger than the {\it GALEX} astrometric uncertainties ($\sim$0.59\arcsec). We use the GCAT {\tt MAG\_FUV} and {\tt MAG\_NUV} magnitudes, which were determined using the SExtractor {\tt AUTO} magnitudes. These magnitudes represent the total galaxy light, and are comparable to the SDSS\, {\tt modelmag} magnitudes \citep{Al.2004}. 

In addition to the total magnitudes, we also retrieve the center 3\arcsec\ magnitudes for the bluest bands (referred to as $NUV_c$ and $u_c$ hereafter). We retrieve {\it WISE} data in the four IR bands using the SDSS DR12 - {\it WISE} matched catalog. We use the {\tt wxmpro} profile-fit magnitudes, and transform them to AB magnitudes to match the convention of the other photometric data. 

We exclude galaxies not detected in $NUV$ or by {\it WISE} from the parent sample. Given the difficulty of reliably modeling the galaxy photometry for large samples, we do not attempt to fill in missing photometric bands via extrapolation from the existing data. The resulting parent sample contains $\sim4\times10^5$ galaxies.

We identify 19,514 ``quiescent Balmer-strong" galaxies (QBS), with Lick H$\delta_{\rm A} > 1.3$ \AA\ in absorption, and H$\alpha$ EQW $< 5$ \AA\ in emission from the parent sample. They have experienced either a recent burst or recent truncation of their star formation. 13,592 of these galaxies also have {\it GALEX} and {\it WISE} photometry. We select 1683 post-starburst (PSB) galaxies with Lick H$\delta_{\rm A} > $ 4 \AA\ and H$\alpha$ EQW $< 3$ \AA\ from the parent sample. These galaxies, a subset of the QBS sample, have less ambiguous star formation histories. 1299 of these galaxies also have {\it GALEX} and {\it WISE} photometry.

The {\it GALEX}, SDSS, and {\it WISE} photometry, the H$\alpha$ EQW and Lick H$\delta_{\rm A}$ measurements, and galaxy coordinates for the whole quiescent Balmer-strong and post-starburst samples are shown in Tables \ref{table:spec} and \ref{table:spec_psb}, including spectroscopically-selected galaxies missing {\it GALEX} or {\it WISE} photometry.

\begin{table}
\begin{rotatetable*}
\begin{deluxetable*}{r r r r r r r r r r r r r r r r r}
\tabletypesize{\scriptsize}
\tablewidth{0pt}
\tablecolumns{17}
\tablecaption{Spectroscopically-Identified Quiescent Balmer-Strong Galaxies from SDSS Parent Training and Test Samples (excerpt) \label{table:spec}}
\tablehead{\colhead{RA} & \colhead{Dec} & \colhead{objid} & \colhead{H$\alpha$ EW} & \colhead{Lick H$\delta_{\rm A}$} & \colhead{NUV} & \colhead{$u$} & \colhead{$g$} & \colhead{$r$} & \colhead{$i$} & \colhead{$z$} & \colhead{W1} & \colhead{W2} & \colhead{W3} & \colhead{W4} & \colhead{NUV$_{c}$} & \colhead{u$_{c}$} \\
\colhead{(deg)} & \colhead{(deg)}  & \colhead{} &\colhead{(\AA)}  & \colhead{(\AA)}  & \colhead{(mag)}  & \colhead{(mag)}  & \colhead{(mag)}  & \colhead{(mag)}  & \colhead{(mag)}  & \colhead{(mag)}  & \colhead{(mag)}  & \colhead{(mag)}  & \colhead{(mag)} & \colhead{(mag)} & \colhead{(mag)} & \colhead{(mag)}} 
\startdata
51.483587 & 1.2720132 &  1237645879578460160 & 4.25 & 5.77 & 20.58 & 18.72 & 17.39 & 16.81 & 16.52 & 16.3 & 16.81 & 17.32 & 17.52 & 15.09 & 19.74 & 22.97 \\
58.023167 & 0.048083 &  1237645942905897216 & 3.99 & 1.32 & 22.35 & 20.4 & 18.5 & 17.34 & 16.71 & 16.2 & 16.67 & 17.13 & 16.17 & 15.54 & 22.36 & 25.76 \\
56.51791 & 0.88506 &  1237645943978983680 & 3.43 & 4.93 & 19.91 & 19.04 & 17.41 & 16.7 & 16.31 & 16.03 & 16.34 & 16.79 & 15.79 & 14.92 & 20.87 & 23.35 \\
245.4847717 & -0.860941 &  1237648672922206464 & 3.75 & 3.64 & 0.0 & 20.02 & 17.97 & 17.07 & 18.94 & 16.13 & 0.0 & 0.0 & 0.0 & 0.0 & 0.0 & 21.65 \\
239.4634552 & -0.4546354 &  1237648673456456192 & 0.3 & 1.5 & 0.0 & 21.57 & 19.05 & 17.82 & 17.31 & 16.88 & 0.0 & 0.0 & 0.0 & 0.0 & 0.0 & 22.64 \\
239.4629364 & -0.0614123 &  1237648673993327104 & 1.28 & 2.48 & 0.0 & 20.63 & 18.94 & 17.99 & 17.58 & 17.17 & 0.0 & 0.0 & 0.0 & 0.0 & 0.0 & 21.4 \\
242.6807556 & -0.143616 &  1237648673994768384 & 3.28 & 1.47 & 0.0 & 19.56 & 17.78 & 16.82 & 16.38 & 16.04 & 0.0 & 0.0 & 0.0 & 0.0 & 0.0 & 20.73 \\
246.62013 & -0.0202279 &  1237648673996473088 & 3.85 & 1.42 & 20.26 & 19.45 & 17.75 & 16.85 & 16.41 & 16.05 & 16.76 & 17.3 & 16.7 & 15.52 & 21.35 & 23.29 \\
239.1622467 & 0.3511362 &  1237648674530066688 & 4.87 & 2.45 & 0.0 & 17.82 & 16.37 & 15.61 & 15.22 & 14.87 & 0.0 & 0.0 & 0.0 & 0.0 & 0.0 & 21.02 \\
241.7365265 & 0.712541 &  1237648675068051968 & 4.33 & 2.27 & 0.0 & 19.03 & 17.05 & 16.05 & 15.5 & 15.08 & 0.0 & 0.0 & 0.0 & 0.0 & 0.0 & 21.4 \\

\enddata
\vspace{0.5cm}
\tablecomments{Table truncated after 10 rows, full table of 9,514 galaxies available online. All magnitudes AB. Missing values denoted with 0.0 values.}
\end{deluxetable*}
\end{rotatetable*}
\end{table}

\begin{table}
\begin{rotatetable*}
\begin{deluxetable*}{r r r r r r r r r r r r r r r r r}
\tabletypesize{\scriptsize}
\tablewidth{0pt}
\tablecolumns{17}
\tablecaption{Spectroscopically-Identified Post-Starburst Galaxies from SDSS Parent Training and Test Samples (excerpt) \label{table:spec_psb}}
\tablehead{\colhead{RA} & \colhead{Dec} & \colhead{objid} & \colhead{H$\alpha$ EW} & \colhead{Lick H$\delta_{\rm A}$} & \colhead{NUV} & \colhead{$u$} & \colhead{$g$} & \colhead{$r$} & \colhead{$i$} & \colhead{$z$} & \colhead{W1} & \colhead{W2} & \colhead{W3} & \colhead{W4} & \colhead{NUV$_{c}$} & \colhead{u$_{c}$} \\
\colhead{(deg)} & \colhead{(deg)}  & \colhead{} &\colhead{(\AA)}  & \colhead{(\AA)}  & \colhead{(mag)}  & \colhead{(mag)}  & \colhead{(mag)}  & \colhead{(mag)}  & \colhead{(mag)}  & \colhead{(mag)}  & \colhead{(mag)}  & \colhead{(mag)}  & \colhead{(mag)} & \colhead{(mag)} & \colhead{(mag)} & \colhead{(mag)}} 
\startdata
199.3198853 & -1.1731747 &  1237648702973411584 & 1.73 & 5.37 & 0.0 & 20.77 & 18.68 & 17.58 & 17.15 & 16.86 & 0.0 & 0.0 & 0.0 & 0.0 & 0.0 & 21.66 \\
203.5754547 & -0.665733 &  1237648703512183040 & 1.43 & 5.65 & 0.0 & 21.11 & 19.14 & 18.15 & 17.77 & 17.49 & 0.0 & 0.0 & 0.0 & 0.0 & 0.0 & 21.81 \\
233.5228729 & -0.7014663 &  1237648703525290240 & 2.9 & 7.16 & 0.0 & 20.74 & 18.72 & 17.83 & 17.31 & 17.03 & 0.0 & 0.0 & 0.0 & 0.0 & 0.0 & 21.28 \\
186.8272858 & -0.4079243 &  1237648704041713664 & 0.96 & 6.02 & 0.0 & 19.34 & 17.68 & 17.0 & 16.72 & 16.47 & 0.0 & 0.0 & 0.0 & 0.0 & 0.0 & 20.22 \\
188.9318 & -0.2237805 &  1237648704042631168 & 1.27 & 6.03 & 18.97 & 17.54 & 16.21 & 15.74 & 15.51 & 15.31 & 16.71 & 17.35 & 16.78 & 15.19 & 22.2 & 19.55 \\
189.06711 & -0.3861852 &  1237648704042696960 & 1.82 & 7.27 & 21.8 & 19.92 & 18.08 & 17.34 & 16.99 & 16.69 & 17.16 & 17.67 & 15.69 & 15.12 & 24.42 & 20.89 \\
203.46138 & -0.271578 &  1237648704048988416 & 1.06 & 5.84 & 0.0 & 19.86 & 17.98 & 17.06 & 16.68 & 16.35 & 0.0 & 0.0 & 0.0 & 0.0 & 0.0 & 21.1 \\
209.1315918 & -0.2689472 &  1237648704051478784 & 1.71 & 5.45 & 0.0 & 19.98 & 18.24 & 17.37 & 17.04 & 16.78 & 0.0 & 0.0 & 0.0 & 0.0 & 0.0 & 20.86 \\
209.24638 & -0.3425346 &  1237648704051544320 & 0.83 & 4.78 & 21.67 & 19.19 & 17.41 & 16.64 & 16.29 & 15.99 & 16.6 & 17.05 & 17.9 & 15.66 & 23.77 & 20.35 \\
213.48737 & -0.308054 &  1237648704053379328 & 0.9 & 4.19 & 22.47 & 20.27 & 18.53 & 17.7 & 17.37 & 17.12 & 17.29 & 17.65 & 17.29 & 15.5 & 24.86 & 20.95 \\

\enddata
\vspace{0.5cm}
\tablecomments{Table truncated after 10 rows, full table of 1683 galaxies available online. All magnitudes AB. Missing values denoted with 0.0 values.}
\end{deluxetable*}
\end{rotatetable*}
\end{table}

\subsection{Modeling LSST Data at Different Redshifts}
\label{sec:models}

\begin{figure}
\includegraphics[width = 0.5\textwidth]{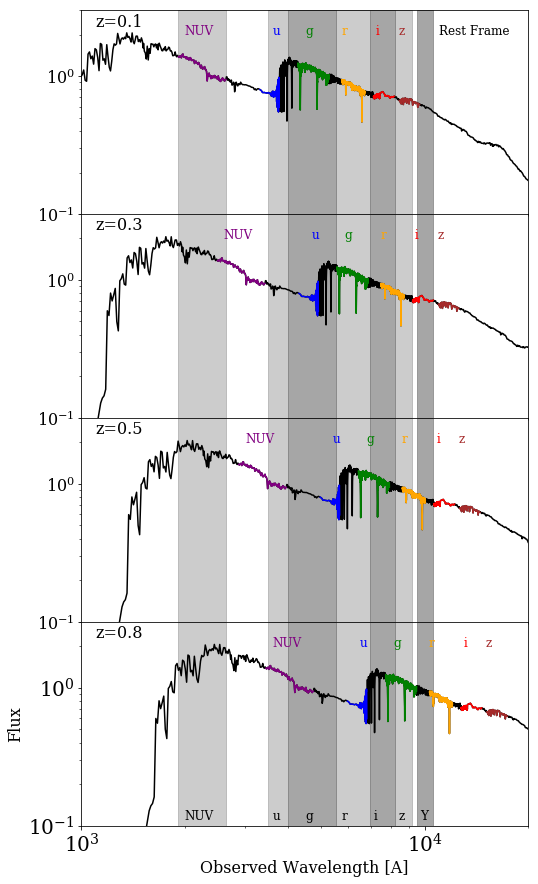}
\caption{Example spectrum of local post-starburst galaxy, and redshifted to $z=0.3, 0.5, 0.8$. The rest-frame $GALEX$ $NUV$ and SDSS-band $ugriz$ portions of the spectrum are colored to demonstrate how they redshift through the LSST $ugrizY$ optical filters (shown shaded in grey) in the observed frame. By redshift $z\sim0.5$, the rest-frame $NUV$ has redshifted into the optical $u$ bands, so this information will be accessible even without deeper {\it GALEX} photometry. {\it WISE} photometry for the 3.4 $\mu$m, 4.6 $\mu$m, 12 $\mu$m, and 22 $\mu$m bands is not shown.}
\label{fig:specsample}
\end{figure}

To model how well we can identify typical TDE host galaxies using photometry available during the LSST era, we use the low redshift ($z\sim0.1$) parent sample described above, and consider what rest-frame observations will be accessible for four redshifts:  $z\sim0.1, 0.3, 0.5, 0.8$. Based on the typical quiescent Balmer-strong and post-starburst galaxy brightnesses in the local sample, and depths obtained by LSST of m$_r<27.5$ and m$_u<26.1$ (5$\sigma$), we expect detections in the $r$ band out to $z\sim0.8$ and in the $u$ band out to $z\sim0.5$.

While most optical/UV bright TDEs have been found at low redshifts \citep[e.g.,][]{vanvelzen2014, Holoien2016}, LSST will probe a much deeper volume than past surveys. Some have predicted that the TDE rate will drop significantly with redshift between $z=$ 0 and 1 \citep{Kochanek2016}, but the combination of a negative k-correction from their extremely blue continuum \citep{Cenko2016}, as well as the rising fraction of post-starburst hosts with redshift \citep{Yan2009,Snyder2011,Wild2016}, may counteract this trend. A full analysis of the expected redshift distribution of the TDEs detectable with LSST is outside the scope of this paper. Instead, we estimate the typical redshift distribution assuming the TDE rate per volume does not change with redshift. For a typical peak brightness of $M_r$ = -19, and a depth for detection of m$_r$=24.5 from a single LSST visit, we will be able to detect TDEs out to $z\sim0.8$. Half of the volume will be contained within $z\lesssim 0.6$, so we use this value for a typical TDE redshift.

The four redshifts we consider are more accurately redshift ranges, because we do not k-correct the colors to a single redshift, to avoid introducing model-dependence. We examine this assumption in \S\ref{sec:kcorr}. For the local sample, the 95$^{\rm th}$ percentile redshift range covers $\Delta z = 0.18$. Our modeled higher redshift ranges are then $z\sim0.3$: $0.24-0.42$, $z\sim0.5$: $0.44-0.62$ (including the typical TDE redshift), and $z\sim0.8$: $0.74-0.92$. Thus, the furthest redshift range considered is comparable to that at which we expect both TDEs and their host galaxies to be detectable.

We use the SDSS and {\it GALEX} photometric data to model the expected LSST photometry. LSST will probe significantly deeper to $r\sim27.5$ mag compared to the SDSS limit of $r\sim22.2$ mag. We assume no evolution in the galaxy properties and that LSST will be sensitive to similar galaxies as the current SDSS survey, but to greater distances. We do not attempt to  transform the magnitudes from the SDSS filters to the LSST filters to avoid introducing model-dependence into the machine learning classifier. We examine these assumptions in \S\ref{sec:kcorr} and \ref{sec:lsst_error}. Instead, we treat the SDSS and {\it GALEX}  photometry as representative of the rest-frame photometry that will be available from LSST for higher redshift galaxies. 

In Figure \ref{fig:specsample} we plot an example post-starburst spectrum generated using FSPS \citep[Flexible Stellar Population Synthesis;][]{Conroy2009, Conroy2010}, at the four modeled redshifts. By $z\sim0.5$, the effective wavelength of the {\it GALEX} {\it NUV} band ($\lambda=2271$ \AA) redshifts to the SDSS $u$ band ($\lambda=3540$ \AA), so LSST $u$ band data can be modeled by the {\it GALEX} {\it NUV} data of analog low redshift galaxies. This is essential to selecting post-starburst and other quiescent Balmer-strong galaxies at higher redshifts, as the {\it NUV} data provides useful information about the young stellar populations. At an intermediate redshift of $z\sim0.3$, the rest-frame $NUV$ will be too faint to be see in the {\it GALEX} All Sky survey, but will not yet have redshifted into the observable $u$ band. The redder optical bands will be unavailable for the higher redshift sample, as this part of the spectrum redshifts past the LSST $Y$ band. We summarize the data we use to model each redshift in Table \ref{table:selection}.

\begin{figure}
\includegraphics[width = 0.5\textwidth]{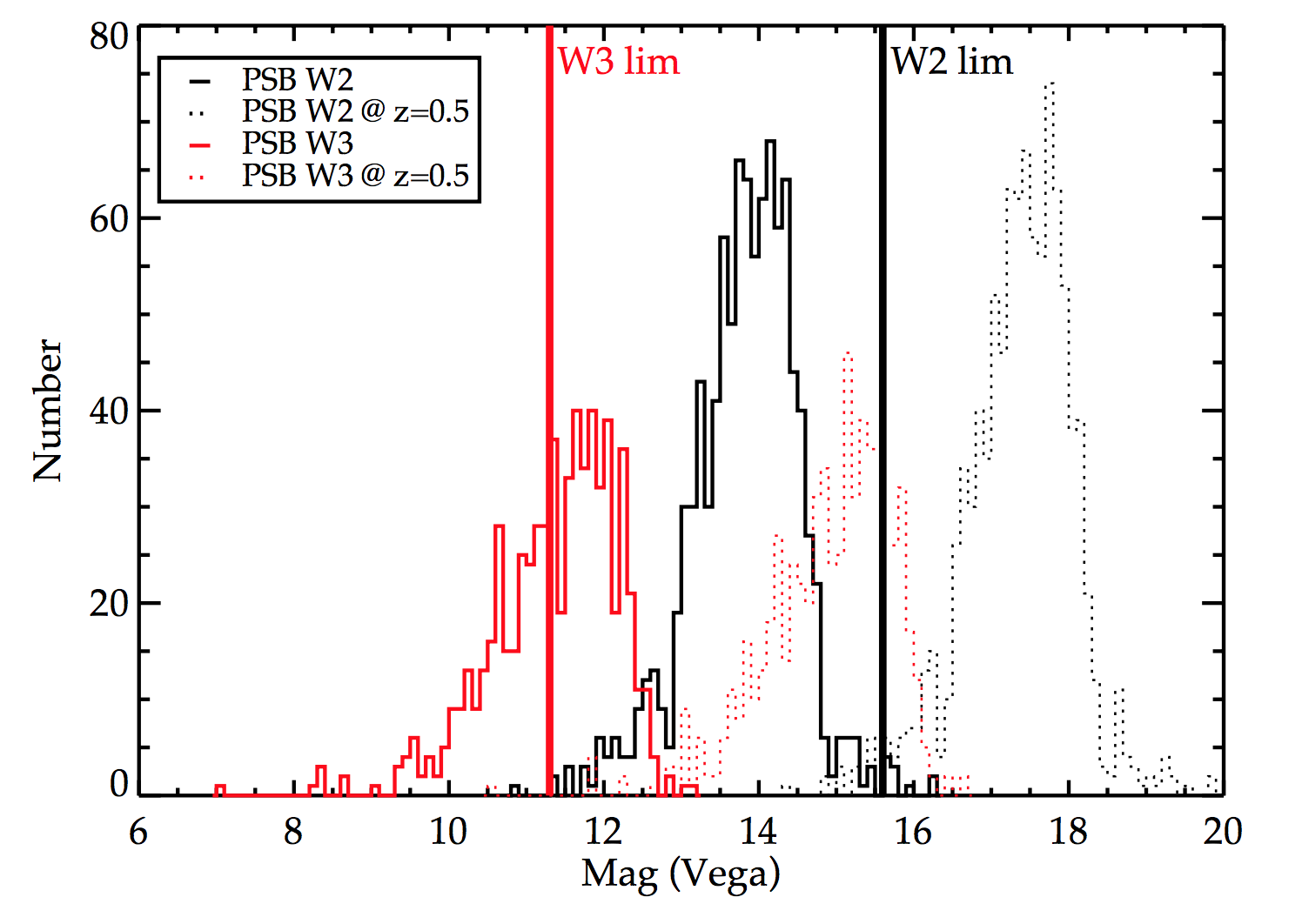}
\caption{{\it WISE} W2 and W3 magnitudes for post-starburst galaxies with SNR$>3$ in these bands. Overplotted vertical lines show the 5$\sigma$ detection limits in the {\it WISE} All-Sky Data Relsease. Dotted histograms show expected distributions of these {\it WISE} magnitudes for post-starburst galaxies at $z=0.5$. While the {\it WISE} W2--W3 color is useful in photometrically classifying post-starburst galaxies, it will only be available to the lower redshift ($z<0.2$) galaxies found with LSST.}
\label{fig:wise}
\end{figure}

In addition to the LSST photometry, {\it GALEX} and {\it WISE} photometry will be available for low redshift (z$\sim0.1$) galaxies. In Figure \ref{fig:wise}, we show the distributions of apparent magnitudes for the local post-starburst sample. For the 12 and 22$\mu$m bands, many galaxies in even the local sample remain undetected. At $z\sim0.5$, the 3.4 and 4.6 $\mu$m bands will also be unavailable. 

In addition to the total galaxy photometry, we also consider the central optical colors using the central 3\arcsec\ photometry from SDSS. For the median redshift of the training sample of $z=0.1$, this corresponds to 5.5 kpc. At $z=0.3$, this spatial scale corresponds to 1.2\arcsec, dropping to 0.90\arcsec\ at $z=0.5$, 0.82\arcsec\ at $z=0.6$, and 0.75\arcsec\ at $z=0.8$. The 75th percentile seeing at LSST is expected to be 0.81\arcsec, so we set $z=0.5$ as the max redshift range where this level of color profile information may be available with LSST data. At the typical redshift expected for LSST to detect TDEs of $z\sim0.6$, all of the bands used for the $z\sim0.5$ redshift range described in Table \ref{table:selection} will be available.

These four redshift ranges are sufficiently large that even coarse estimates of the photometric redshift from LSST will be accurate enough to place the galaxy into one of these redshift ranges. \citet{Salvato2018} predict the accuracy of LSST's photometric redshifts to be $\sigma_z/(1+z) \sim 0.025$. Thus, the following discussion of applying our host galaxy selection to LSST assumes that we will know at least the photometric redshift in advance to accuracies easily achievable with LSST photometry.

\begin{table}
\centering
\caption{Rest-frame photometry available in LSST era}
\label{table:selection}
\begin{tabular}{r r r r}
\hline
\hline
$z\sim0.1$ &  $z\sim0.3$ & $z\sim0.5$ & $z\sim 0.8$ \\
$NUV-u$  & $u-g$ & $NUV-u$ & $NUV-u$ \\
$u-g$ & $g-r$ & $u-g$ & $u-g$ \\
$g-r$ & $r-i$ & $g-r$ &  \\
$r-i$ & $ u - u_{c}$ & $ u - u_{c}$ & \\
$i-z$ &  & $ u - NUV_{c}$ & \\
$ u - u_{c}$ & & \\
$ u - NUV_{c}$ & &\\
$z - [3.4]$ & &\\
$[3.4] - [4.6]$ &  &\\
$[4.6] - [12]$ & & \\
$[12] - [22]$ & &\\

\hline

\end{tabular}
\end{table}

\subsection{Distinctive Features}
\label{sec:distinctivefeatures}

\begin{figure*}
\includegraphics[width = 0.5\textwidth]{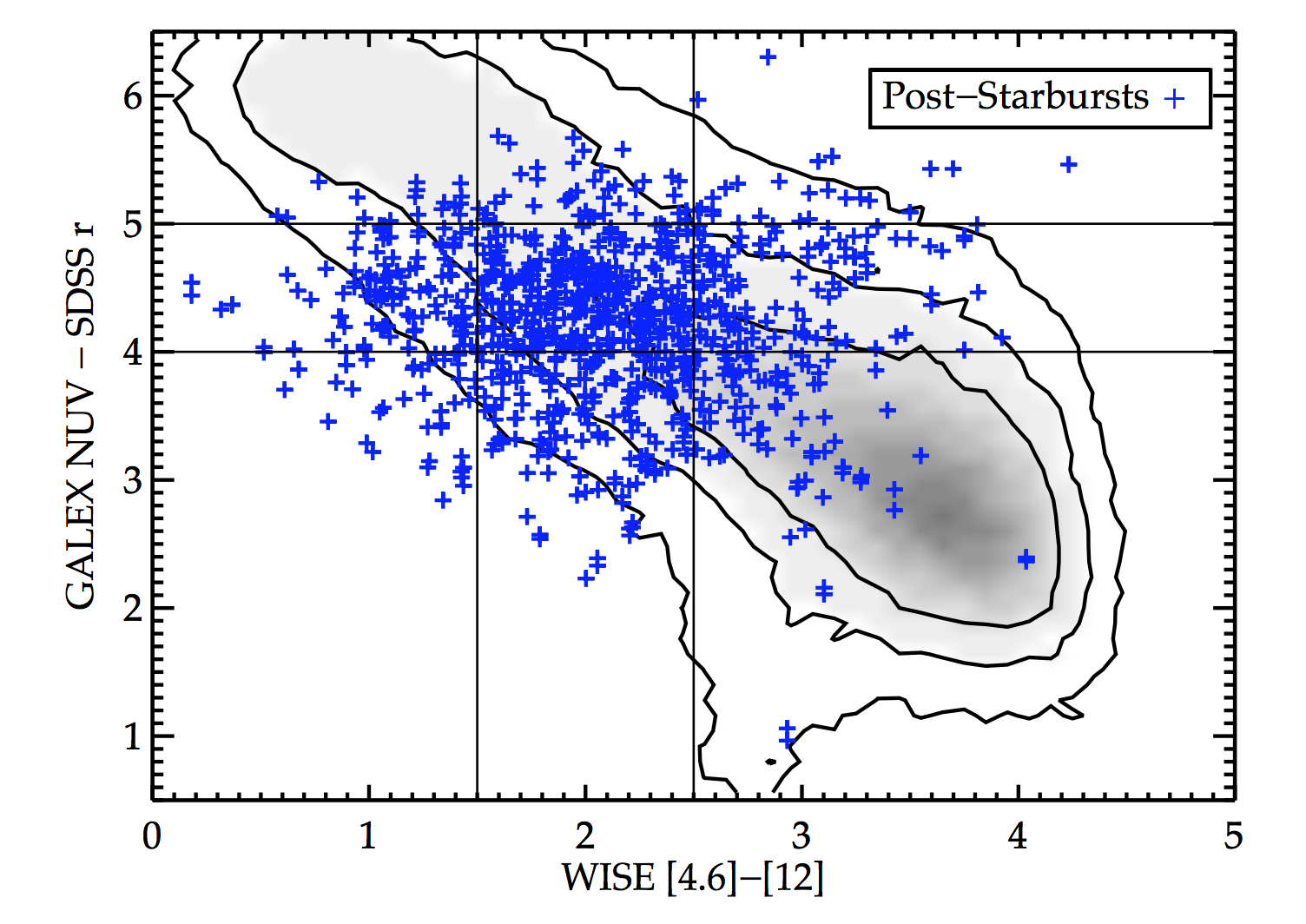}
\includegraphics[width = 0.5\textwidth]{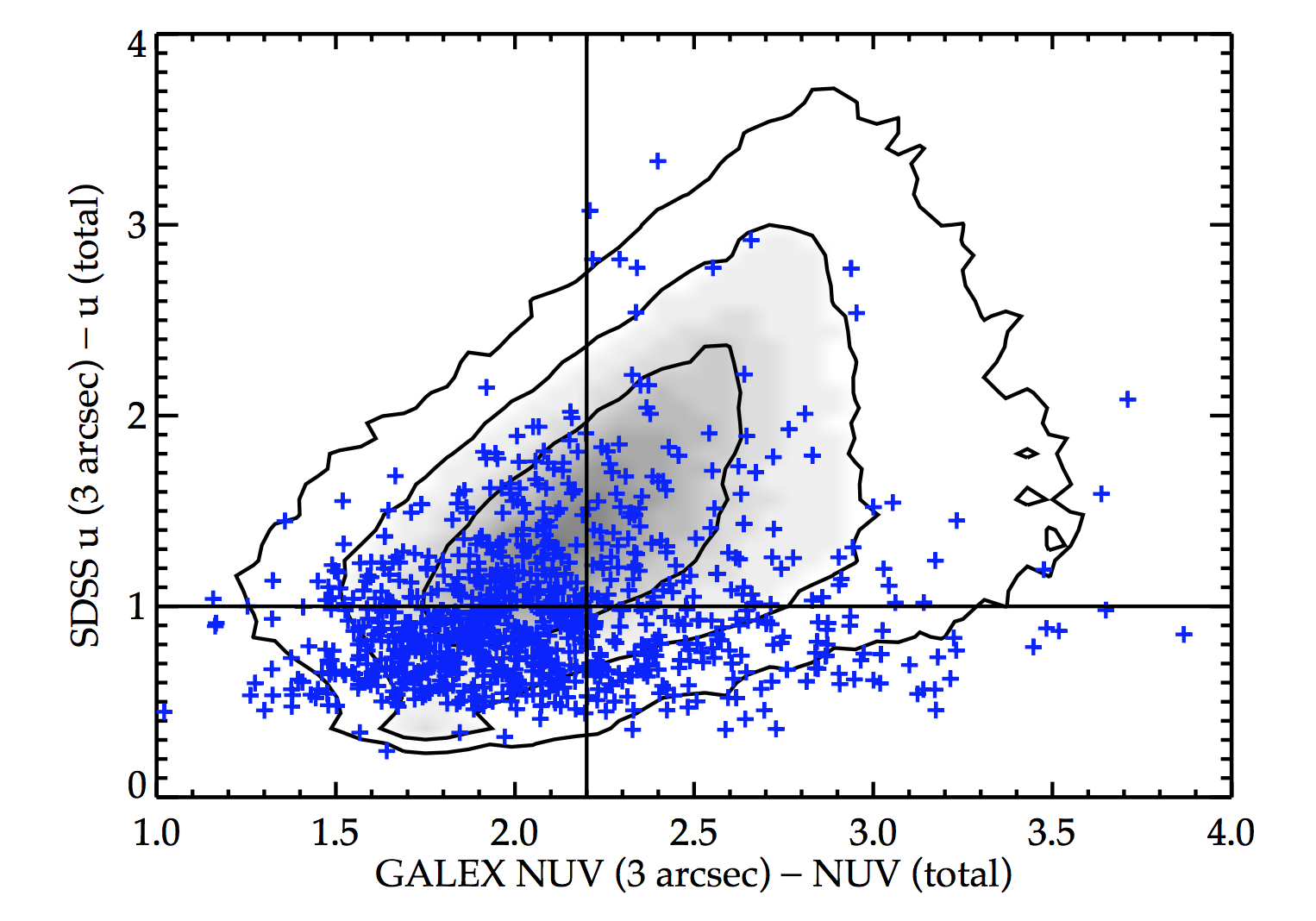}
\caption{Example of simple photometric-only selection of post-starburst galaxies demonstrating the difficulty of achieving high purity samples of favored TDE host galaxies. {\bf Left:} $NUV-r$ color vs. [4.6]--[12]$\mu$m colors for SDSS galaxies (black contours) and post-starburst galaxies (blue points). While post-starburst galaxies have been shown to lie in each green valley \citep{Wyder2007, Wong2012, Schawinski2014,Ko2013, Yesuf2014, Alatalo2014}, they compose a very small fraction of the green valley galaxies. Selecting on the box outlined in black results in a sample with very low purity: 1\% for post-starburst galaxies, and 7\% for quiescent Balmer-strong galaxies.
{\bf Right:} $NUV$ and $u$ band colors of total - center photometry for SDSS galaxies (black contours) and post-starburst galaxies (blue points). Post-starburst galaxies have been shown to have concentrated blue light \citep{Yang2006}, which allows for better selection than the green valley alone. However, adding the simple cuts defined in black here to those in the left-hand figure only increases the purity of the quiescent Balmer-strong selection to 16\%.
}
\label{fig:uvir}
\end{figure*}

Post-starburst galaxies occupy the UV-optical green valley \citep{Wyder2007, Wong2012, Chilingarian2012, Schawinski2014}, as well as the IR green valley \citep{Ko2013, Yesuf2014, Alatalo2014, Alatalo2016c}. The location of the post-starbursts in this color-color space is shown in Figure \ref{fig:uvir}a. A selection in the green valley can be defined to capture the locus of the post-starbursts, but contamination from other galaxies due to dust or different star formation histories limits the use of this selection. The selection indicated in Figure \ref{fig:uvir}a using $NUV-r$ color and [4.6]--[12] $\mu$m color results in a sample with very low purity: 1\% for post-starburst galaxies, and 7\% for quiescent Balmer-strong galaxies.

The fact that many post-starbursts lie in the green valley is not their only unique photometric property. Post-starbursts are also known to have blue cores \citep{Yang2006, Yan2006}, concentrated light \citep{Yang2004, Yang2008}, and bright $NUV$ luminosities. In Figure \ref{fig:uvir}b we plot the [center -- total] colors for the $NUV$ and $u$ band photometry, showing the unique concentrated blue light observed for post-starburst galaxies. However, even including the two blue core selection cuts shown does not efficiently select post-starbursts. Adding these two cuts to the previous green valley selection increases the purity of the quiescent Balmer-strong selection to only 16\%. Such an impure sample would waste valuable follow-up time on misidentified post-starburst galaxies. Thus, another approach is needed, if we are to efficiently select post-starburst galaxies using only photometric data.

\subsection{Random Forest Classification}

While the number of galaxies used here is small ($\sim4\times10^5$) by machine learning standards, this method can nonetheless greatly improve the selection of post-starburst and other quiescent Balmer-strong galaxies over the by-hand selection considered above. The Random Forest method \citep{Breiman2001} can classify objects with high accuracy, using a large number of observable features without risking overfitting the classifier to the specific training sample. Random Forests have been used to classify astrophysical sources such as variable stars and transient events \citep[e.g.,][]{Richards2011,Brink2013} and quasar candidates \citep{Carrasco2015}.

The Random Forest method works by generating a large number of decision trees, which each then ``vote'' on the classification of an object. To generate each tree, a subsample is chosen with replacement from the training sample. A set of random cuts are tested from some portion of the available features. Here, the number of features each tree uses to cut on is the square root of the total number of features available. The cut that maximizes the Gini impurity at each node of the tree is selected, then the tree is grown down to the next level \citep{Breiman1984}. This process continues until there are no possible additional splits, because the sample is all the same class (sometimes because only a single object remains). Because each tree has only considered a subset of the data and of the available selection features, this method prevents overfitting. 

\begin{figure*}
\includegraphics[width=\textwidth]{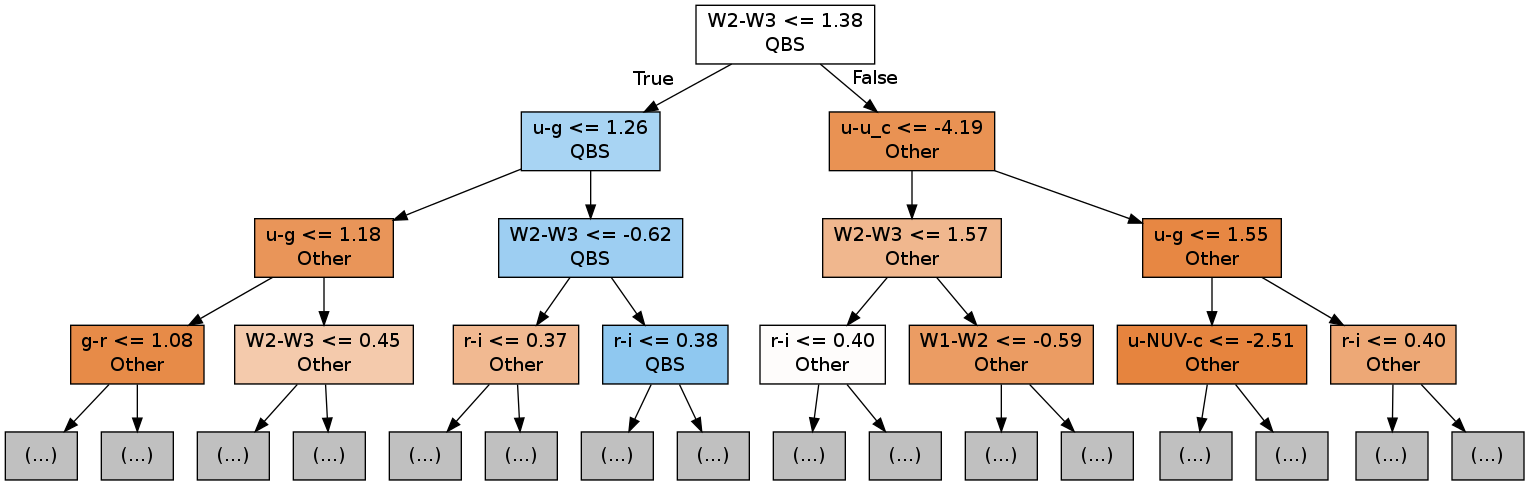}
\caption{Example decision tree from the Random Forest for classifying quiescent Balmer-strong (QBS) galaxies from imaging data alone. To generate each tree, galaxy observable features like color or concentration are chosen randomly, and the cut that maximizes the Gini impurity is chosen at each node. The example tree is truncated after 3 levels, but each tree in the Random Forest is grown until all leaves are pure, consisting of only one class. The Random Forest can then be used to classify galaxies, with each decision tree casting a vote on the galaxy classification. In this example, cuts are shown to distinguish QBS hosts from other galaxies. Blue nodes contain a higher fraction of QBS galaxies, and orange nodes contain a higher fraction of non-QBS galaxies.}
\label{fig:tree}
\end{figure*}

We grow 50 trees for the Random Forest classifier, and note that the results do not change qualitatively beyond 10 trees. We make use of the Random Forest classifier form the {\tt scikit-learn} package \citep{scikit-learn}. An example of an individual tree is shown in Figure \ref{fig:tree}.

We split our initial parent sample of SDSS galaxies (\S\ref{sec:data}) into a training sample and test sample, each with 203,046 galaxies. We use the training sample to construct the Random Forest. We are interested in identifying both quiescent Balmer-strong galaxies and their subset of post-starburst galaxies as described above. We train the Random Forest to identify each class from the rest of the sample. The SDSS spectra are used to label each galaxy as quiescent Balmer-strong or other, and in separate training, post-starburst or other.

Once the Random Forest has been grown using the training sample, we use the test sample to assess how well it performs. To test the Random Forest, each galaxy in the test sample is classified. The fifty trees like the one shown in Figure \ref{fig:tree} are used to classify each galaxy as quiescent Balmer-strong or other. Each tree then ``votes" on the likely classification of the galaxy, and the winning class is assigned by majority vote. 

We compare the actual classification from the SDSS spectra to the predicted classification from the Random Forest. We present the purity and completeness of the post-starburst (or quiescent Balmer-strong) selections to demonstrate the performance of the method. We also test several additional machine learning classifiers on this problem, using {\tt scikit-learn} packages to perform k-nearest neighbors \citep[KNN;][]{Altman1992}, Support Vector Machine \citep[SVM;][]{Smola2004}, Adaboost \citep{Hastie2009}, and Gradient Boosting \citep{Friedman2001} classifiers. Random Forest provides the best combination of purity and completeness for this application; even when the other methods have similar completeness, the purity is much lower.

\subsection{Feature Selection}

\begin{figure}
\includegraphics[width = 0.5\textwidth]{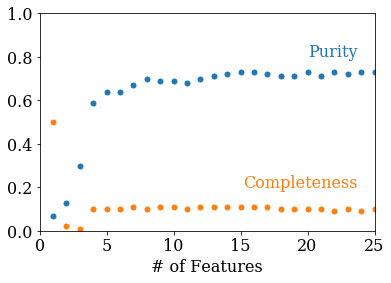}
\caption{Purity and completeness of selecting quiescent Balmer-strong galaxies starting with the NUV to W4 apparent magnitudes, sequential colors, measures of color profile information, and signal-to-noise ratios---the total feature set that would be available at $z\sim0.1$ in the LSST era---and iteratively removing the feature with the lowest importance. The change in purity is gradual, until fewer than four features are left. Because most of the first features to be eliminated are the magnitudes and signal-to-noise ratios, we choose a final feature set consisting of only colors (Table \ref{table:selection}) to avoid adding noise into the classification. In \S\ref{sec:results}, we explore the possibilities at higher redshifts where fewer data will be available, and examine the ranked feature importances to gain an understanding of which features are the most useful.}
\label{fig:pur_comp}
\end{figure}

An important consideration for using the Random Forest method is on which observable features to select. Why not use every combination available of observed colors and magnitudes? There is disagreement in the literature as to whether noisy features hurt the classifier performance \citep{Biau2010, Brink2013}. We use two tests to determine which features are best to use, out of a total feature set of SDSS, {\it GALEX}, and {\it WISE} apparent magnitudes, the 3\arcsec\ $u$ and $NUV$ magnitudes, as well as the sequential colors $NUV-u$ through W3-W4. We also consider the difference between the central and total $u$ and $NUV$ magnitudes, and the signal-to-noise ratios of the NUV and W4 magnitudes. We do not attempt to k-correct the colors of the various samples, to avoid introducing model-dependent effects. This decision is discussed further in \S\ref{sec:kcorr}.

The first test is similar to that used by \citet{DIsanto2016}. We first use the total feature set, construct the Random Forest, and iteratively remove the least important feature. The purity and completeness for each number of features in this process is shown in Figure \ref{fig:pur_comp}. The change in purity is gradual, until fewer than four features are left. The trends in this analysis are due to a combination of (1) the Random Forest's increasing effectiveness with enough features from which to select multiple subsets, and (2) the number of features that will encode sufficient information for high purity classification. We explore the second idea further in \S\ref{sec:results}. The improvement in purity, even for a small number of features, over the 16\% purity achieved using the by-hand cuts in \S\ref{sec:distinctivefeatures} is due to the optimization in cuts provided by the Random Forest method.

As an alternative to this iterative method, we test excluding all features with lower importances than a feature containing randomly generated numbers. If a real feature is less important in constructing the Random Forest than a random variable, it is likely adding noise instead of helping the classification. Thus, we construct a Random Forest including the random variable, then re-generate the Random Forest excluding all features determined less important than the random variable \citep[e.g.,][]{Brink2013}. The purity and completeness that result are comparable to the plateau seen in the method described above and in Figure \ref{fig:pur_comp}. 

We also test whether sequential colors or colors skipping one or more features in between perform best. Sequential colors produce the highest purity and completeness. This is likely due to the low importance of features spanning a large wavelength range, e.g., $z$-W1. This feature has a low importance, and a feature set that includes multiple such colors (e.g., $i$-W1 and $z$-W2) will have more unimportant features.

In the iterative method, most of the first features to be eliminated are the apparent magnitudes and SNRs, so we choose a feature set consisting of only colors. The advantage of colors over apparent magnitudes is likely due to the removal of the first-order dependence on distance. The list of features is shown in Table \ref{table:selection}.

\section{Results}
\subsection{Photometric Selection Method Performance}
\label{sec:results}

\begin{figure}
\includegraphics[width = 0.5\textwidth]{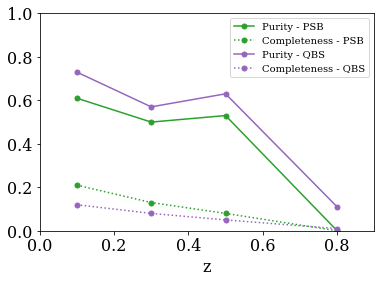}
\caption{Purity and completeness for the quiescent Balmer-strong (QBS) sample and its post-starburst galaxy (PSB) subsample, using training data whose rest-frame equivalents are or will be available in the LSST era at various redshifts. As redshift increases, features are lost as the data becomes unavailable and as rest frame data redshifts out of the observed bands; conversely, new data becomes available as the rest frame data redshifts into the observed bands, in particular the rest frame {\it NUV} from z=0.3 to z=0.5. The performance of the method changes both with the number of features (see Table \ref{table:selection}) and {\it which} features are available. Both of these effects are seen from z=0.1 to z=0.3 and 0.5. By z=0.8, the method is primarily limited by the small number of features available. We explore which features are most important in \S\ref{sec:results} and Figure \ref{fig:features}.}
\label{fig:pc}
\end{figure}

The sample purity and completeness derived from this method are displayed in Table \ref{table:results}. Due to the random nature of the process, the uncertainties in these numbers are $\sim1$ percentage point. The Random Forest classifier successfully selects more than $>60$\% of the spectroscopically-confirmed quiescent Balmer-strong or post-starburst samples from our test sample of $2\times10^5$ galaxies. These high purities are significant given the low intrinsic rates of these galaxies in the test sample.

We plot the purity and completeness versus redshift in Figure \ref{fig:pc}. While the best performance comes from the low redshift sample with the full feature set, a small but significant increase in purity is seen from $z\sim 0.3-0.5$ as the rest-frame $NUV$ data redshifts into the $u$ band. The completenesses for the low-redshift sample are 10--20\%. A higher completeness is obtained for the post-starburst sample than for the broader quiescent Balmer-strong sample, perhaps due to the greater range of recent star formation histories of the galaxies in the latter sample. 

\begin{table}
\centering
\caption{Random Forest Classification}
\label{table:results}
\begin{tabular}{r r r}
\hline
\hline
 & Quiescent Balmer-Strong & Post-Starburst \\
$z\sim0.1$ & P = 61\%, C = 21\% & P = 73\%, C = 12\% \\
$z\sim0.3$ & P = 50\%, C = 13\% & P = 57\%, C = 8\% \\
$z\sim0.5$ & P = 53\%, C = 8\% & P = 63\%, C = 5\% \\
$z\sim0.8$ & P = 11\%, C = 1\% & P = 0\%, C = 0\% \\
\hline
\end{tabular}
Purity (P) and completeness (C) of galaxies selected using our Random Forest classifier.
\end{table}

\begin{figure*}
\includegraphics[width = 0.48\textwidth]{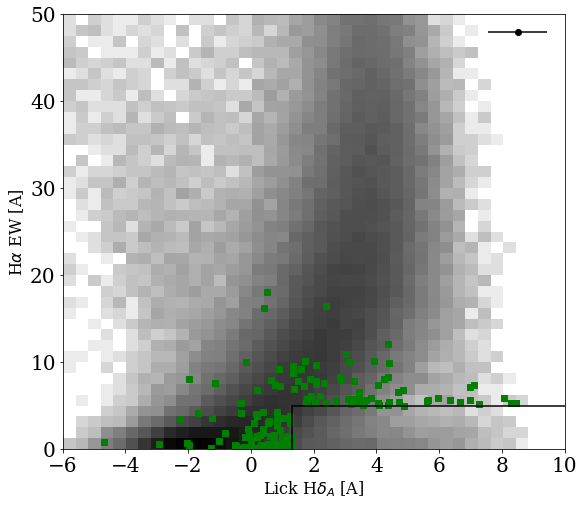}
\includegraphics[width = 0.48\textwidth]{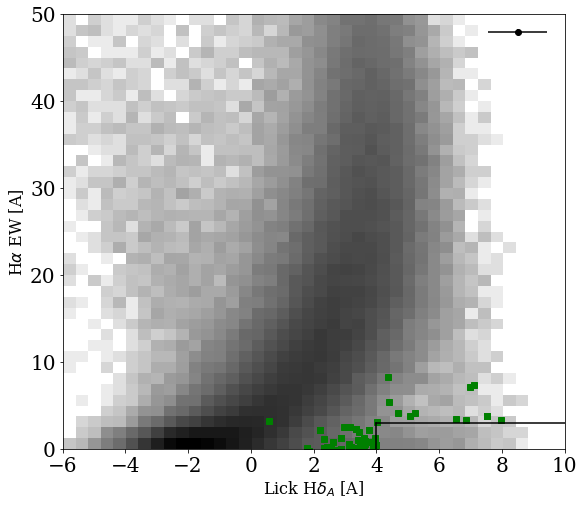}
\caption{``Contaminating" galaxies: H$\alpha$ EW emission and Lick H$\delta_A$ absorption of galaxies that were selected photometrically to be quiescent Balmer-strong (left) or post-starburst (right). The parent distribution is shown in greyscale. Characteristic errorbars are shown in the upper right of each panel. Because these false-positives are clustered near the edge of the selection cuts in each case, the effective purity of our photometric-only selection method is likely underestimated, as the contaminating galaxies may not have significantly different physical properties.}
\label{fig:contam_hahd}
\end{figure*}

The purity is high ($>50$\%) out to $z\sim0.5$, but the lack of available features at $z\sim0.8$ shows the limits of this method. We note that the purity may be underestimated, as the ``contaminating" galaxies are clustered just outside the spectroscopic selection for both the post-starburst and quiescent Balmer-strong samples. We plot the H$\alpha$ EW emission and Lick H$\delta_A$ absorption of these galaxies in Figure \ref{fig:contam_hahd}. Most of the ``contaminating" false-positives in the post-starburst sample would in fact satisfy the quiescent Balmer-strong criteria. Thus, the contaminating galaxies may not be significantly different in their physical properties that the post-starburst galaxies within our spectroscopic cuts.

\begin{figure*}
\includegraphics[width = 0.9\textwidth]{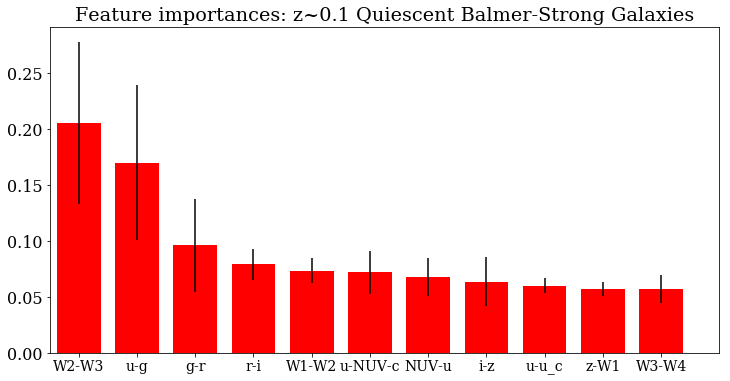}
\includegraphics[width = 0.9\textwidth]{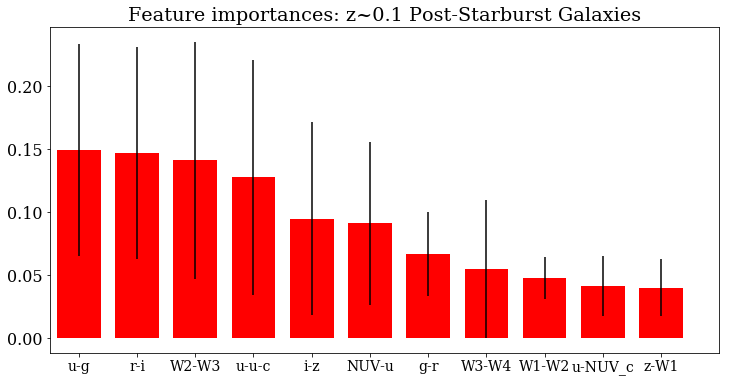}
\caption{Feature importances (Gini importance) as determined by the Random Forest classification for our training sample of quiescent Balmer-strong and post-starburst galaxies. The most useful host galaxy observable features are optical $u-g$ and $r-i$ colors and the {\it WISE} [4.6]--[12]$\mu$m color.}
\label{fig:features}
\end{figure*}

The feature importances shown in Figure \ref{fig:features} allow us to inspect how the Random Forest classifications work. For the quiescent Balmer-strong sample, the most important features are the [4.6]--[12] $\mu m$ and $u-g$ colors. For the post-starburst sample, the rankings are similar to the quiescent Balmer-strong sample, given the large errors caused by the smaller sample size. The primary difference we observe between the quiescent Balmer-strong and post-starburst samples is the increased importance of the profile information for the post-starburst galaxies, especially at higher redshifts when fewer features are available. We found in \citet{French2017} that the main physical difference in galaxies selected by relaxing the H$\delta$ selection cut was to find galaxies with less stellar mass produced in the recent starbursts. If the quiescent Balmer-strong galaxies had less extreme starbursts, the centrally concentrated blue light may be a less distinguishing feature for them. The $z-$W1 feature is consistently ranked low, and reflects the differing portions of the spectrum probed by the $z$ and W1 bands, sensitive primarily to the stellar populations and dust, respectively.

Many of the most useful features, such as the {\it WISE} photometry, are unavailable at higher redshift and lead to the decrease in performance with redshift.  However, the most useful features, especially $u-g$,  are robust with redshift, even as other features drop out. From our feature selection analysis in Figure \ref{fig:features} and described above, the greatest difference in performance comes from the four most important features. However, the ranked feature importance for the quiescent Balmer-strong sample shows only two dominant features, with the rest nearly equally important given the errors. We test whether a Random Forest classifier constructed with only four features depends on what the third and fourth feature are. While the feature importances are similar, the method shows higher purity and completeness when higher ranked features such as $g-r$ and $r-i$ are used instead of lower ranked features such as $z-$W1 and W3--W4.

Both the purity and completeness of the photometric classification are lower for the higher redshifts than the $z\sim0.1$ sample. However, $<30$\% contamination is still quite low. LSST is expected to find 1000s of TDEs per year \citep{vanvelzen2011}, so even $\sim10$\% completeness will yield 100s of TDEs per year. The decrease in performance is caused by both a decrease in the number of features and the loss of useful features. Features are lost and gained with redshift as the rest-frame data redshifts through the available observed bands. In particular, the rest frame {\it NUV} is gained from z=0.3 to z=0.5. However, there is a net loss in features with redshift as the galaxies become too faint for {\it GALEX} and {\it WISE} photometry, and too small for accurate spatially resolved photometry. By z=0.8, the method is primarily limited by the small amount of features available. 

\subsection{Discovering TDEs with LSST using their Host Galaxies}

\citet{vanvelzen2011} predict 4131 TDEs will be discoverable with LSST every year, scaled from the SDSS Stripe 82 detection rate. For the sample of TDE candidates with broad H/He lines considered in \citet{French2016}, 75\% are in quiescent Balmer-strong galaxies. This fraction is lower in other somewhat different samples of TDE candidates, e.g., \citet{Graur2017} find 36\% of TDEs have quiescent Balmer-strong hosts\footnote{using the definition in \S\ref{sec:data}}. We thus assume that a range of 36-75\% of TDEs will occur in quiescent Balmer-strong hosts. We assume a completeness of 8\% using the typical TDE redshift as described in \S\ref{sec:models}. Following up all transient events that occur in photometrically selected quiescent Balmer-strong galaxies would then yield 119 - 248 real TDEs per year. This will dramatically improve the sample of known TDEs from its current rate of a few per year. In the next section, we explore the efficiency of following up photometrically-selected quiescent Balmer-strong galaxies, given contaminants from the false-positives selected and from other transient events likely in quiescent Balmer-strong and post-starburst galaxies.

Even before LSST, this strategy will help identify TDEs in ZTF. ZTF is expected to find $\sim30$ TDEs per year. Assuming again that 36-75\% of TDEs will occur in quiescent Balmer-strong hosts, and assuming 21\% completeness given the lower redshift of likely ZTF events, following up all transient events in photometrically selected quiescent Balmer-strong galaxies would yield 2-4 TDEs per year.

\subsection{Likelihood for Transients in Quiescent Balmer-Strong Hosts to be TDEs}
\label{sec:othertransients}

In order for selecting quiescent Balmer-strong galaxies to be a feasible way to identify TDEs, the TDE rate must dominate over other transient events. Because these galaxies are quiescent, the supernova rates are quite low. Using the supernova rates per unit mass from \citet{Mannucci2005}, for a $1\times10^{10}$ stellar mass E/S0 galaxy, the rates are $4\times10^{-4}$ per year for type Ia, $<9\times10^{-5}$ per year for types Ib/c, and $<1\times10^{-4}$ per year for type II supernovae, for a total of  $<6\times10^{-4}$ per year. The TDE rate is between $2\times10^{-4}$ - $3\times10^{-3}$ per year in the quiescent Balmer-strong and post-starburst samples \citep{French2016}. 

Thus, the TDE rate will be $>0.3-5\times$ the supernova rate, given the range of likely TDE rates, especially for post-starburst galaxies, which appear to have a higher TDE rate than quiescent Balmer-strong galaxies. Following up all transients in quiescent Balmer-strong galaxies will yield a sample of candidate events that are 25--83\% real TDEs, assuming a perfect host selection. Assuming a host galaxy selection purity of 53\%, following up all transients in photometrically-selected host galaxies will yield a sample of 13--42\% real TDEs. This assumption about the host galaxy selection purity may be an underestimate of the true purity of likely TDE host galaxies, given the clustering of our ``false positives" near the selection boundary (Figure \ref{fig:contam_hahd}). We thus estimate 13--83\% of transients followed up will be real TDEs, accounting for the TDE rates, supernova rates, and host galaxy selection purity. Therefore, following up all transients will yield 119--248 TDEs as discussed above, as well as 24--1700 supernovae per year.

As a comparison, a method selecting on transients in normal quiescent galaxies would have a similar supernova rate, but much lower TDE rates of $\sim 10^{-6} - 10^{-4}$ per galaxy per year. A sample of transients in quiescent galaxies thus would yield a sample of 0.2--16\% TDEs, significantly worse than using photometrically-selected quiescent Balmer-strong galaxies. Follow-up then would require significantly more time, resulting in follow-up of 600--120,000 supernovae per year to detect the same 119--248 TDEs.

The quiescent Balmer-strong and post-starburst galaxies are selected to be quiescent, and as a result, to have already experienced the supernovae of the O and B stars formed in the recent starburst. However, they still contain younger stellar populations than most E/S0 galaxies. Even if the total supernova rate (of types Ia, II, and Ib/c combined) was 10$\times$ higher than the rate expected by \citet{Mannucci2005} for the E/S0 sample, and more consistent with the rate for the S0a/b sample, a rate of $1\times10^{-3}$ per year is still comparable to the expected TDE rate in these galaxies. 

We note that the time period where the TDE rate is higher than the supernova rate may extend further back in time towards the starburst; several predictions \citep{Madigan2017, Stone2017} for the TDE rate enhancement predict it to be a declining function with time since the starburst.

The ability to locate a transient event to the nucleus will further improve the likelihood that any given transient is a TDE. However, both post-starburst galaxies \citep{Yang2008} and TDE hosts \citep{Graur2017,Law-Smith2017} have centrally concentrated young stellar populations, suggesting that most contaminating supernovae also will be centrally concentrated. Most newly formed stars in these galaxies will be within $\lesssim4$ kpc \citep[e.g.,][]{Swinbank2012}. The relative astrometry for LSST is predicted to be $\sim 0.02$\arcsec\ \citep{lsstscireq}. At the redshifts of most TDEs, this corresponds to 0.1 kpc. 

Assuming an effective radius of 1 kpc \citep[appropriate for $z\sim0.5$ early type galaxies;][]{vanderwel2014}, and a stellar mass profile where density ($\rho$) scales with radius ($r$) as $\rho \propto r^{-2}$, we thus expect to find 1/20${\rm th}$ of the stellar mass within the 0.1 kpc distinguishable by LSST. This greatly reduces the contamination from supernovae. In quiescent Balmer-strong galaxies, the TDE rate would be 6--100$\times$ greater than the supernova rate, with 46--99\% of followed-up objects being TDEs and the number of contaminating supernovae only 3--140 per year. In contrast, for early type galaxies, the TDE rate would be 0.03--3$\times$ the supernova rate, with 3--77\% of followed-up objects being TDEs, and the number of contaminating supernovae 74--3.8$\times10^3$ per year.

Further candidate events will be filtered at later times using subsequent light curve data or spectroscopic follow-up.

AGN variability is unlikely to significantly contaminate the predicted transients in quiescent Balmer-strong galaxies. \citet{vanvelzen2011} find a low probability ($<2\times10^{-5}$) of a TDE-like flare occurring after a long period of inactivity in variable AGN. AGN show more variability in the X-ray than optical \citep{Ulrich1997}, but even X-ray flares in galaxies not known to be active are thought to be TDEs rather than unusual AGN \citep{Donley2002}. However, the AGN flare rate in quiescent Balmer-strong galaxies is not well studied; Pan-STARRS and ZTF lightcurves of quiescent Balmer-strong galaxies will be able to address this issue before LSST.

\begin{deluxetable*}{r r r r r r r r r r r}
\tabletypesize{\scriptsize}
\tablewidth{0pt}
\tablecolumns{10}
\tablecaption{Pan-STARRS + {\it WISE} Newly-Identified Quiescent Balmer-Strong Galaxies (excerpt) \label{table:phot_ps}}
\tablehead{\colhead{RA} & \colhead{Dec} & \colhead{objID} &  \colhead{$g$} & \colhead{$r$} & \colhead{$i$} & \colhead{$z$} & \colhead{W1} & \colhead{W2} & \colhead{W3} \\
\colhead{(deg)} & \colhead{(deg)}   & \colhead{} & \colhead{(mag)}  & \colhead{(mag)}  & \colhead{(mag)}  & \colhead{(mag)}  & \colhead{(mag)}  & \colhead{(mag)}  & \colhead{(mag)}} 
\startdata
20.9056474 & -16.6176169 &    88050209055919232 & 17.62 & 16.83 & 16.49 & 16.26 & 16.52 & 17.05 & 17.53 \\
159.5967173 & -16.6227059 &    88051595967573120 & 19.41 & 18.61 & 18.19 & 17.97 & 17.82 & 18.14 & 17.48 \\
336.2176631 & -16.6448188 &    88023362176636608 & 18.08 & 17.37 & 17.02 & 16.81 & 16.85 & 17.20 & 17.14 \\
18.3322384 & -16.5943124 &    88080183321867104 & 20.27 & 19.98 & 19.42 & 19.24 & 18.81 & 19.11 & 17.27 \\
346.8757987 & -16.8567198 &    87773468757592320 & 19.84 & 18.65 & 18.16 & 17.95 & 17.53 & 17.88 & 17.04 \\
197.9699045 & -16.5985086 &    88081979699742112 & 21.26 & 20.2 & 19.81 & 19.44 & 19.06 & 19.10 & 17.52 \\
350.0856371 & -16.6223928 &    88053500856443568 & 21.46 & 20.88 & 20.38 & 20.26 & 19.71 & 19.80 & 17.5 \\
325.1038088 & -16.6241404 &    88053251037941392 & 19.83 & 19.15 & 18.88 & 18.58 & 19.12 & 20.05 & 17.73 \\
188.8527772 & -16.5961736 &    88081888529424992 & 20.01 & 19.63 & 19.35 & 19.22 & 19.51 & 20.12 & 17.95 \\
13.5798407 & -16.5915556 &    88090135797820496 & 19.09 & 18.46 & 18.14 & 17.98 & 17.88 & 18.27 & 16.77 \\

\enddata
\tablecomments{Table truncated after 10 rows, full table of 57299 galaxies available online. All magnitudes AB. griz magnitudes in Pan-STARRS system.}
\end{deluxetable*}

\begin{deluxetable*}{r r r r r r r r r r r}
\tabletypesize{\scriptsize}
\tablewidth{0pt}
\tablecolumns{10}
\tablecaption{DES + {\it WISE} Newly-Identified Quiescent Balmer-Strong Galaxies (excerpt) \label{table:phot_des}}
\tablehead{\colhead{RA} & \colhead{Dec} & \colhead{coadd\_object\_id} & \colhead{$g$} & \colhead{$r$} & \colhead{$i$} & \colhead{$z$} & \colhead{W1} & \colhead{W2} & \colhead{W3} \\
\colhead{(deg)} & \colhead{(deg)}  & \colhead{} & \colhead{(mag)}  & \colhead{(mag)}  & \colhead{(mag)}  & \colhead{(mag)}  & \colhead{(mag)}  & \colhead{(mag)}  & \colhead{(mag)}} 
\startdata
26.162398 & -5.10442 & 266816340 & 17.67 & 16.90 & 16.57 & 16.36 & 16.86 & 17.35 & 16.73 \\
24.826391 & -4.817912 & 235013753 & 16.54 & 15.73 & 15.40 & 15.15 & 15.64 & 16.18 & 15.44 \\
25.381688 & -4.583149 & 243264243 & 18.45 & 17.55 & 17.19 & 16.95 & 17.22 & 17.5 & 17.11 \\
25.8201 & -4.77617 & 266797394 & 18.39 & 17.57 & 17.20 & 16.95 & 17.29 & 17.70 & 17.11 \\
25.674348 & -3.972962 & 251236728 & 18.15 & 17.37 & 17.0 & 16.76 & 17.09 & 17.56 & 16.73 \\
25.968743 & -3.527678 & 268924270 & 17.29 & 16.54 & 16.21 & 15.97 & 16.43 & 16.93 & 16.07 \\
26.056814 & -3.580794 & 268927243 & 17.02 & 16.34 & 16.04 & 15.81 & 16.36 & 16.86 & 16.98 \\
21.27813 & -2.760255 & 221490749 & 18.79 & 17.97 & 17.62 & 17.37 & 17.72 & 18.07 & 17.32 \\
20.774063 & -1.676664 & 216898857 & 17.14 & 16.39 & 16.06 & 15.84 & 16.35 & 16.89 & 16.26 \\
21.531674 & -2.318308 & 221600179 & 17.93 & 17.19 & 16.85 & 16.61 & 17.01 & 17.45 & 17.02 \\

\enddata
\tablecomments{Table truncated after 10 rows, full table of 9337 galaxies available online. All magnitudes AB. griz magnitudes in DES system.}
\end{deluxetable*}

\begin{deluxetable*}{r r r r r r r r r r r}
\tabletypesize{\scriptsize}
\tablewidth{0pt}
\tablecolumns{11}
\tablecaption{SDSS + {\it WISE} Newly-Identified Quiescent Balmer-Strong Galaxies (excerpt) \label{table:phot}}
\tablehead{\colhead{RA} & \colhead{Dec} &  \colhead{objid} & \colhead{$u$} & \colhead{$g$} & \colhead{$r$} & \colhead{$i$} & \colhead{$z$} & \colhead{W1} & \colhead{W2} & \colhead{W3} \\
\colhead{(deg)} & \colhead{(deg)}   & \colhead{} & \colhead{(mag)}  & \colhead{(mag)}  & \colhead{(mag)}  & \colhead{(mag)}  & \colhead{(mag)}  & \colhead{(mag)}  & \colhead{(mag)}  & \colhead{(mag)}} 
\startdata
116.764104 & -0.8339698 &  1237646795988140288 & 21.11 & 19.24 & 18.34 & 17.96 & 17.68 & 17.79 & 18.19 & 17.2 \\
197.9930265 & -1.1741177 &  1237648702972887040 & 20.38 & 18.44 & 17.62 & 17.26 & 17.04 & 16.93 & 17.33 & 16.5 \\
189.9654787 & -0.6628777 &  1237648703506219264 & 20.76 & 19.04 & 18.02 & 17.67 & 17.44 & 17.47 & 17.86 & 17.11 \\
229.3670079 & -0.776028 &  1237648703523455488 & 20.5 & 18.77 & 18.0 & 17.68 & 17.4 & 17.6 & 18.05 & 17.61 \\
205.0925687 & -0.7082981 &  1237648703512838400 & 20.84 & 18.97 & 18.15 & 17.82 & 17.58 & 17.41 & 17.74 & 17.46 \\
217.4976164 & -0.2464468 &  1237648704055148800 & 20.97 & 19.3 & 18.17 & 17.8 & 17.58 & 17.49 & 17.88 & 17.42 \\
225.1479114 & 0.1013984 &  1237648704595362048 & 21.1 & 19.33 & 18.36 & 17.92 & 17.59 & 17.86 & 18.29 & 17.79 \\
155.3279148 & -0.9076529 &  1237648720146202880 & 20.44 & 18.71 & 17.95 & 17.63 & 17.39 & 17.62 & 17.72 & 17.05 \\
229.5090881 & 0.2887919 &  1237648721789256704 & 21.12 & 19.41 & 18.57 & 18.21 & 17.98 & 18.24 & 18.44 & 17.63 \\
200.3267334 & 0.6757371 &  1237648722313347328 & 20.87 & 19.23 & 18.36 & 17.98 & 17.74 & 17.76 & 18.04 & 16.92 \\

\enddata
\tablecomments{Table truncated after 10 rows, full table of 848 galaxies available online. All magnitudes AB. ugriz magnitudes in SDSS system.}
\end{deluxetable*}

%%%%%%%%%%%

\begin{deluxetable*}{r r r r r r r r r r r}
\tabletypesize{\scriptsize}
\tablewidth{0pt}
\tablecolumns{10}
\tablecaption{Pan-STARRS + {\it WISE} Newly-Identified Post-starburst Galaxies (excerpt) \label{table:phot_ps_psb}}
\tablehead{\colhead{RA} & \colhead{Dec} & \colhead{objID} &  \colhead{$g$} & \colhead{$r$} & \colhead{$i$} & \colhead{$z$} & \colhead{W1} & \colhead{W2} & \colhead{W3} \\
\colhead{(deg)} & \colhead{(deg)}   & \colhead{} & \colhead{(mag)}  & \colhead{(mag)}  & \colhead{(mag)}  & \colhead{(mag)}  & \colhead{(mag)}  & \colhead{(mag)}  & \colhead{(mag)}} 
\startdata
336.2176631 & -16.6448188 &    88023362176636608 & 18.08 & 17.37 & 17.02 & 16.81 & 16.85 & 17.20 & 17.14 \\
342.3582229 & -16.6675957 &    87993423582379264 & 21.49 & 19.99 & 19.42 & 19.11 & 18.15 & 18.68 & 17.79 \\
193.7133521 & -16.7667089 &    87881937133970432 & 19.97 & 19.5 & 19.1 & 18.96 & 19.04 & 19.04 & 17.57 \\
27.060048 & -16.7621758 &    87880270599735744 & 18.55 & 17.49 & 17.03 & 16.85 & 16.88 & 17.36 & 17.31 \\
157.5626928 & -16.7353274 &    87911575626297936 & 19.66 & 18.86 & 18.25 & 18.13 & 17.62 & 17.71 & 16.0 \\
336.322675 & -16.6734522 &    87993363226542224 & 19.42 & 18.72 & 18.34 & 18.17 & 17.92 & 18.22 & 16.20 \\
98.6205694 & -16.588656 &    88090986205293968 & 20.89 & 20.22 & 19.63 & 19.19 & 18.21 & 18.54 & 16.47 \\
302.2295861 & -16.715314 &    87943022295922000 & 22.05 & 21.13 & 21.13 & 20.57 & 18.85 & 18.75 & 16.75 \\
351.4121364 & -16.5968016 &    88083514121594176 & 19.97 & 18.87 & 18.33 & 18.13 & 17.64 & 18.02 & 16.93 \\
227.9810852 & -16.6105995 &    88062279810287728 & 21.37 & 19.75 & 19.23 & 19.01 & 17.97 & 18.38 & 17.48 \\

\enddata
\tablecomments{Table truncated after 10 rows, full table of 9690 galaxies available online. All magnitudes AB. {\it griz} magnitudes in Pan-STARRS system.}
\end{deluxetable*}

\begin{deluxetable*}{r r r r r r r r r r r}
\tabletypesize{\scriptsize}
\tablewidth{0pt}
\tablecolumns{10}
\tablecaption{DES + {\it WISE} Newly-Identified Post-starburst Galaxies (excerpt) \label{table:phot_des_psb}}
\tablehead{\colhead{RA} & \colhead{Dec} & \colhead{coadd\_object\_id} & \colhead{$g$} & \colhead{$r$} & \colhead{$i$} & \colhead{$z$} & \colhead{W1} & \colhead{W2} & \colhead{W3} \\
\colhead{(deg)} & \colhead{(deg)}  & \colhead{} & \colhead{(mag)}  & \colhead{(mag)}  & \colhead{(mag)}  & \colhead{(mag)}  & \colhead{(mag)}  & \colhead{(mag)}  & \colhead{(mag)}} 
\startdata
22.512555 & -2.947905 & 224821207 & 18.51 & 17.56 & 17.27 & 17.06 & 17.13 & 17.48 & 17.32 \\
25.515892 & -0.595226 & 249988895 & 17.79 & 17.11 & 16.80 & 16.61 & 17.05 & 17.53 & 16.82 \\
28.118299 & -22.853303 & 258853250 & 18.81 & 17.87 & 17.51 & 17.29 & 17.56 & 17.80 & 17.02 \\
28.569581 & -21.061182 & 259693519 & 17.95 & 17.15 & 16.85 & 16.62 & 16.94 & 17.27 & 16.61 \\
30.727404 & -18.87177 & 68521246. & 17.38 & 16.68 & 16.38 & 16.18 & 16.5 & 17.01 & 16.97 \\
32.567998 & -22.677123 & 83101028. & 18.63 & 17.76 & 17.42 & 17.20 & 17.5 & 17.82 & 17.38 \\
36.79265 & -21.774958 & 119194150 & 17.22 & 16.5 & 16.20 & 15.97 & 16.38 & 16.87 & 16.39 \\
34.696696 & -19.113486 & 97477464. & 18.92 & 18.14 & 17.82 & 17.62 & 17.87 & 18.13 & 17.36 \\
28.085827 & -15.60385 & 257249250 & 17.72 & 17.02 & 16.71 & 16.5 & 16.94 & 17.44 & 16.88 \\
30.287858 & -16.491598 & 153096709 & 19.06 & 18.20 & 17.86 & 17.65 & 17.87 & 18.20 & 17.39 \\

\enddata
\tablecomments{Table truncated after 10 rows, full table of 753 galaxies available online. All magnitudes AB. {\it griz} magnitudes in DES system.}
\end{deluxetable*}

\begin{deluxetable*}{r r r r r r r r r r r}
\tabletypesize{\scriptsize}
\tablewidth{0pt}
\tablecolumns{11}
\tablecaption{SDSS + {\it WISE} Newly-Identified Post-starburst Galaxies (excerpt) \label{table:phot_psb}}
\tablehead{\colhead{RA} & \colhead{Dec} &  \colhead{objid} & \colhead{$u$} & \colhead{$g$} & \colhead{$r$} & \colhead{$i$} & \colhead{$z$} & \colhead{W1} & \colhead{W2} & \colhead{W3} \\
\colhead{(deg)} & \colhead{(deg)}   & \colhead{} & \colhead{(mag)}  & \colhead{(mag)}  & \colhead{(mag)}  & \colhead{(mag)}  & \colhead{(mag)}  & \colhead{(mag)}  & \colhead{(mag)}  & \colhead{(mag)}} 
\startdata
229.3670079 & -0.776028 &  1237648703523455488 & 20.5 & 18.77 & 18.0 & 17.68 & 17.4 & 17.6 & 18.05 & 17.61 \\
259.3197541 & 58.2280318 &  1237651226246119680 & 21.68 & 19.95 & 18.91 & 18.55 & 18.35 & 18.43 & 18.73 & 17.9 \\
229.7708601 & 56.6946225 &  1237651250972066048 & 20.57 & 18.91 & 17.89 & 17.52 & 17.3 & 17.45 & 17.72 & 17.58 \\
265.284451 & 19.7025259 &  1237651250991924224 & 20.81 & 18.9 & 17.94 & 17.63 & 17.42 & 17.55 & 17.86 & 17.51 \\
209.0260394 & 66.426363 &  1237651539790135296 & 20.47 & 18.79 & 17.81 & 17.46 & 17.25 & 17.44 & 17.83 & 17.98 \\
131.554487 & 52.7311131 &  1237651191895163136 & 20.53 & 18.7 & 17.79 & 17.43 & 17.19 & 17.22 & 17.54 & 17.53 \\
183.6934534 & 1.3400826 &  1237651752401305600 & 20.31 & 18.66 & 17.97 & 17.71 & 17.57 & 17.65 & 17.91 & 17.25 \\
242.8176741 & 53.3829864 &  1237651539799310592 & 20.95 & 19.26 & 18.5 & 18.2 & 17.97 & 18.13 & 18.64 & 18.54 \\
204.2425535 & 65.542822 &  1237651271904002304 & 20.53 & 18.75 & 17.97 & 17.66 & 17.41 & 17.52 & 17.89 & 17.67 \\
171.5635778 & 65.8063978 &  1237651271361298688 & 20.25 & 18.48 & 17.75 & 17.47 & 17.2 & 17.58 & 17.97 & 17.48 \\

\enddata
\tablecomments{Table truncated after 10 rows, full table of 117 galaxies available online. All magnitudes AB. {\it ugriz} magnitudes in SDSS system.}
\end{deluxetable*}

\subsection{New Catalogs of Quiescent Balmer-Strong and Post-starburst Galaxies}
\label{sec:newcat}

Having trained the Random Forest classifier using spectroscopically-confirmed quiescent Balmer-strong galaxies from the SDSS, we can now identify new quiescent Balmer-strong and post-starburst galaxies without accompanying spectroscopy. We apply our machine learning model to galaxies in the large photometric surveys Pan-STARRS DR1 \citep{Chambers2016, Flewelling2016} and Dark Energy Survey DR1 \citep[DES][]{Abbott2018} cross-matched with {\it WISE}, as well as to galaxies in the SDSS but too faint to have spectroscopy. We use the same training and test samples described above, with quiescent Balmer-strong and post-starburst galaxies spectroscopically identified from the SDSS spectra. For each photometric survey, we limit and modify the SDSS training sample photometry to match the rest-frame data of that survey.

Because these new photometric surveys are deeper than the SDSS training data, the associated {\it GALEX} and {\it WISE} magnitudes are fainter. We eliminate consideration of features that have low signal-to-noise. Thus, we remove the {\it GALEX NUV}, {\it WISE} 22$\mu$m photometry, and the 3\arcsec\ aperture photometric measurements. We re-train the Random Forest classifier without these features, using the sample with SDSS spectra described above to train and test the classifier. If we only include the SDSS $ugriz$ photometry and {\it WISE} W1, W2, and W3 band photometry, the expected performance is a purity of 71\% and completeness of 12\% for the quiescent Balmer-strong galaxies and a purity of 61\% and completeness of 20\% for the post-starburst galaxies. Neither Pan-STARRS nor DES has $u$-band photometry, so for these samples we similarly retrain the Random Forest using only SDSS $griz$ photometry and {\it WISE} W1, W2, and W3 band photometry. The expected performance in this case is 69\% purity and 11\% completeness for the quiescent Balmer-strong galaxies and 61\% purity and 20\% completeness for the post-starburst galaxies. This is not significantly worse than for the larger set of measurements used in \S\ref{sec:results} (see Table \ref{table:results}).

Because of the {\it WISE} data requirement, the galaxies considered here are limited in redshift compared to the full DES, PanSTARRS, and SDSS samples (see Figure \ref{fig:wise}). Thus, we do not attempt to use {\it a priori} knowledge of the redshifts of galaxies in the new samples, like we plan to do for LSST. We consider the effects of the differing redshift and absolute magnitude distributions between the SDSS training sample and the new photometric-only samples in \S\ref{sec:zdep} and \ref{sec:magdep}.

To verify the samples of quiescent Balmer-strong and post-starburst galaxies selected from the photometric-only surveys, we compare below the number obtained to that expected from the rate of quiescent Balmer-strong and post-starburst galaxies (3.3\% and 0.32\%, respectively) from the spectroscopic training/test samples and the estimated completeness of the classifier.

\subsubsection{Pan-STARRS + WISE Photometric-only Catalog of Quiescent Balmer-Strong Galaxies}

We select galaxies from the Pan-STARRS DR1 catalog to classify. We require at least three visits for each of the $griz$ bands and require good photometry with {\tt QfPerfect $\ge$ 0.95} for each band. We separate the stars from the galaxies with the criteria {\tt rMeanPsfMag - rMeanKronMag > 0.5}. The resulting sample contains 30 million galaxies. We cross-match this sample with {\it WISE} photometry, selecting W1, W2, and W3. The resulting catalog contains 22 million galaxies. We use the photometric transformations from \citet{Tonry2012} to convert the SDSS training data to the Pan-STARRS system. 

Using the Random Forest classifier trained with SDSS $griz$ and {\it WISE} W1-W3 photometry described above, we identify 57299 candidate galaxies as quiescent Balmer-strong (Table \ref{table:phot_ps}). Given the initial sample of 21,887,693 Pan-STARRS galaxies, a quiescent Balmer-strong rate of 3.3\% from the SDSS spectroscopic training/test samples described in \S\ref{sec:data}, and an estimated completeness of 11\%, we expect 79K quiescent Balmer-strong galaxies. The actual number of galaxies classified as quiescent Balmer-strong here is 72\% of this number. Thus, the actual performance of the classifier is close to that expected from the spectroscopic SDSS test sample. 

We also identify 9690 candidate galaxies as post-starburst (Table \ref{table:phot_ps_psb}). Given the initial sample of Pan-STARRS galaxies, a post-starburst rate of 0.32\%, and an estimated completeness of 20\%, we expect 14,008 galaxies. We thus select 69\% of the expected sample.

\subsubsection{DES + WISE Photometric-only Catalog of Quiescent Balmer-Strong Galaxies}

We obtain galaxies from the DES DR1 using the {\tt des\_dr1.galaxies} catalog of galaxies and \\ {\tt des\_dr1.des\_allwise} catalog of objects cross-matched to {\it WISE} from the NOAO Data Lab\footnote{\url{http://datalab.noao.edu/desdr1}}. We select galaxies with signal-to-noise ratios of $>3$ in the DES $griz$ bands, and in the {\it WISE} W1, W2, and W3 bands. The resulting sample contains 3,656,666 galaxies. We use the photometric transformations from \citet{Tie2017} to convert the DES data to the SDSS system.

We use the same Random Forest classifier generated for the Pan-STARRS classification described above, with a purity of 69\% and a completeness of 11\%. We identify 9337 candidate galaxies as quiescent Balmer-strong from the DES + {\it WISE} photometric sample (Table \ref{table:phot_des}). 

Given the initial sample of 3.6M galaxies, a quiescent Balmer-strong rate of 3.3\% from the spectroscopic training and test samples described in \S\ref{sec:data}, and an estimated completeness of 11\%, we expect 13K quiescent Balmer-strong galaxies. The actual number of galaxies classified as quiescent Balmer-strong is 70\% of this, similar to the performance seen in the Pan-STARRS sample described above.

We also identify 753 candidate galaxies as post-starburst (Table \ref{table:phot_des_psb}). Given the initial sample of DES galaxies, a post-starburst rate of 0.32\%, and an estimated completeness of 20\%, we expect 2340 galaxies. We thus have selected 32\% of the expected sample, an efficiency less than for the quiescent Balmer-strong galaxies, and less than for the Pan-STARRS catalog.

\subsubsection{SDSS + WISE Photometric-only Catalog of Quiescent Balmer-Strong Galaxies}

There are a large number of galaxies with photometry, but no spectroscopy, in the SDSS. Here we draw a photometric-only sample of galaxies from the SDSS, beyond the cuts made to select the spectroscopic sample \citep{Strauss2002}. We select objects with $r$-band Petrosian magnitudes $17.7 <r <22.2$, fainter than the spectroscopic limit of 17.7 mag, but still detected. We require a modelMag of $u<22$ mag to ensure that the $u$-band is detected. We select only objects without spectra (specobjid=0), clean photometry (clean=1), and likely to be galaxies (type=3). The spectroscopic sample requires surface brightnesses of $\mu_{50} < 24.5$ mag arcsec$^{-2}$ in the $r$ band, and we relax this to 27.5 mag arcsec$^{-2}$. We require the galaxies to have {\it WISE} W1, W2, and W3 measurements. This new sample contains 3.26M galaxies. 

Using the Random Forest classifier trained using SDSS $ugriz$ photometry and {\it WISE} W1-W3 bands, we select an additional 848 quiescent Balmer-strong galaxy candidates (Table \ref{table:phot}). Given the initial sample of 3.26M galaxies, a quiescent Balmer-strong rate of 3.3\% from the spectroscopic training and test samples, and an estimated completeness of 12\%, we expect 12,909 quiescent Balmer-strong galaxies, implying that the performance of the Random Forest classifier is $\sim15\times$ worse than expected and much worse than for the Pan-STARRS and DES surveys. We also identify 117 candidate galaxies as post-starburst (Table \ref{table:phot_psb}). Given  a post-starburst rate of 0.32\% and an estimated completeness of 20\%, we expect 2089 galaxies. We thus select only 6\% of the expected sample. The lower signal-to-noise of the SDSS photometry is the likely source of this poor performance.

With the three new photometric samples presented here, there are 67,484 new candidate quiescent Balmer-strong galaxies and 10,560 new candidate post-starburst galaxies. These new samples increase the numbers of such galaxies by 3.5$\times$ and $6.3\times$, respectively, compared to previously-known spectroscopically identified samples. Given the high probability that a transient in one of these galaxies is a TDE, as discussed in \S\ref{sec:othertransients}, this new catalog can be used to identify likely TDEs for follow up observations at the time of their discovery. This new catalog also will be a boon to galaxy evolution studies.

\section{Discussion}

\subsection{Redshift-Dependent Effects}
\label{sec:zdep}

In this section, we explore the effects of applying a Random Forest model based on the SDSS spectroscopic survey to the LSST, Pan-STARRS, DES, and SDSS photometric surveys, which are deeper, i.e., which extend to higher redshifts and fainter absolute magnitudes. We start here with the redshift effects. We test the different wavelength bands and filter transmission curves. We also test the separate redshift assumptions made for (1) future selection from LSST and (2) the current selection from Pan-STARRS, DES, and SDSS, all cross-matched with {\it WISE} data. Photometric redshifts from LSST photometry will be good enough to place the host galaxies in the four redshift bins discussed above, but for the Pan-STARRS, DES, and photometric SDSS galaxies, the {\it WISE} requirement restricts the redshifts and we use only the $z\sim0.1$ SDSS training/test sample. We test the dependence of our selection purity and completeness on redshift by varying the redshift range of the training/test sample, as well as the effects of using a training sample with a test sample at a different redshift range.

\begin{figure*}
	\includegraphics[width = 0.49\textwidth]{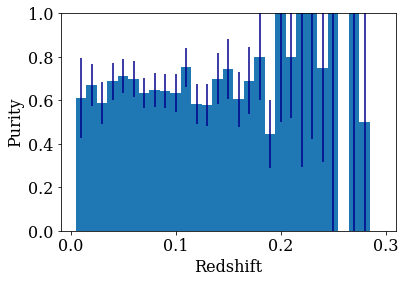}
	\includegraphics[width = 0.49\textwidth]{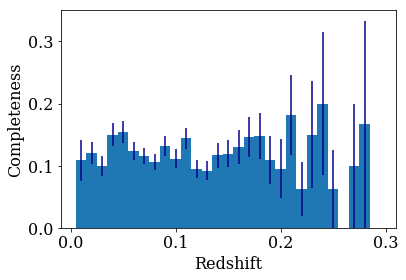}
	\caption{Dependence of purity and completeness on redshift for the quiescent Balmer-strong galaxy selection from the SDSS test sample. There are no significant declines in purity or completeness with redshift, which implies that the selection will not suffer for incrementally higher redshift objects (see \S\ref{sec:zdep}).}
	\label{fig:z_dependence}
\end{figure*}

If our selection method declines in performance with redshift, the selection from deeper surveys would suffer. We test the dependence of the purity and completeness of the SDSS test sample on redshift (Figure \ref{fig:z_dependence}). We find no significant declines in either metric with redshift.

\subsubsection{Redshift Evolution}
We have assumed throughout that there is no evolution between galaxies in the training/test samples and the galaxies we intend to select from other surveys, e.g., that galaxies at $z\sim0.5$ are the same as those at $z\sim0.1$, a difference of $\sim4$ Gyr. Not until $z>1$ do we expect evolution in typical post-starburst galaxies. Such high redshift post-starbursts are fundamentally different, possessing ``A" rather than ``K+A" star spectra. Post-starburst galaxies observed at $z\sim0.5$ had time for a significant old ($5-10$ Gyr) population to have formed before the starburst.

\subsubsection{K-corrections}
\label{sec:kcorr}

Applying k-corrections to the galaxy samples might improve our selection method by making the colors more accurate, but could also harm it if the assumed spectral models are not appropriate or difficult to fit. We first test the theoretical impact of an ideal k-correction by applying our selection method to narrow redshift slices. To make use of the greatest number of galaxies, we select several narrow slices of $\Delta z = 0.02$ near the median $z\sim0.1$. The purity and completeness can rise by $5-8$ percentage points. If photo-$z$s reliable enough to k-correct the colors could be obtained, {\it and one were willing to assume a set of likely spectral models}, the Random Forest results could be improved.

Next, we test the impact of several k-correction methods. As a reference, we compare to the un-corrected $ugriz$ photometry, from which the Random Forest can select quiescent Balmer-strong galaxies with a purity of 61\% and completeness of 7\%. The most recent version of the SDSS {\tt photoz} catalog k-corrected magnitudes \citep[last updated in DR12][]{Beck2016} produces a purity of 6\% and a completeness of 69\%. Using k-corrected magnitudes from an older version of the {\tt photoz} catalog \citep[from DR10, see][]{Csabai2002} results in a purity of 42\% and a completeness of 2\%. The large variation in purity and completeness obtained using different k-correcting methods demonstrates the lack of robustness in training on the k-corrected catalogs. We obtain similar results using the magnitudes k-corrected to either $z=0$ or $z=0.1$. 

We also test the k-correcting method of \citet{Chilingarian2010}, \citet{Chilingarian2011a}, finding a purity of 56\% and a completeness of 6\%. This method allows for the addition of the $NUV$ and central $NUV, u$ photometry. Adding these produces a purity of 69\% and a completeness of 9\%, compared with the purity of 70\% and completeness of 10\% obtained from the same set of un-corrected magnitudes. Thus, while perfect k-corrections would help the selection method, methods for k-correcting large photometric samples can result in worse performance, with the best case result similar to using the un-corrected values.

We compare the galaxies selected with and without k-corrected photometry to see if they are the same. We find a 67\% overlap, of which most (77\%) are true positives. We test whether the false positives are any better in the k-corrected photometry case. The scatter in H$\alpha$ and H$\delta_A$ is similar to the un-k-corrected case, but the median H$\alpha$ emission is higher, i.e., further from the selection criteria. There is thus no clear benefit in using k-corrected magnitudes at this time.

\subsubsection{Are SDSS training redshifts appropriate for LSST?}
\label{sec:lsst_error}

In estimating the performance of the Random Forest classifier trained on SDSS data in \S\ref{sec:results} for LSST data, and in quantifying the decrease in performance with redshift, we have assumed a perfect mapping between {\it GALEX}/SDSS and LSST filters. Here, we consider the effects of different filter transmission curves between {\it GALEX}/SDSS and LSST and the different rest-frame portions of the galaxy spectra they will be sensitive to between $z\sim0.1$ and $z\sim0.6$, the expected redshift typical of LSST TDEs.

In the four redshift bins we consider in \S\ref{sec:results}, the rest frame UV/optical will be accessible through different observed--frame filters. We assume, for example, that the SDSS $u-g$ color at $z\sim0$ models the observed $g-r$ color of the same galaxy at $z=0.6$ observed with LSST, as these bands bracket the 4000 \AA\ break at each redshift. Similarly, we assume that the observed-frame  $NUV-g$ color can model the expected LSST $u-i$ color.

However, given the wide redshift range between $z\sim0.1$ and $0.6$, the differing bandpasses will lead to systematic offsets in $u-g$ and $NUV-g$ of $0.4-0.5$ mag. While the colors are correlated, the Random Forest classifier is not robust to such shifts. Thus, while k-correcting the colors did not improve the classifier performance for the $z\sim0.1$ samples (\S\ref{sec:kcorr}), k-corrections will be necessary for applying our $z\sim0.1$ trained classifier to LSST galaxies.

\begin{figure*}
\centering
\includegraphics[width = 0.425\textwidth]{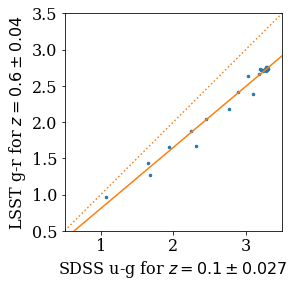}
\includegraphics[width = 0.49\textwidth]{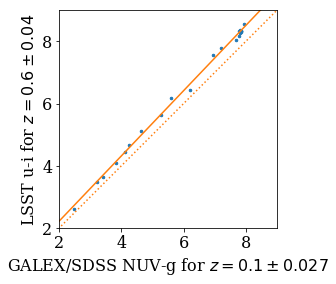}
\caption{Comparison of synthetic photometry for a range of star-forming, quiescent, and post-starburst galaxies. We compare (left) the synthetic SDSS $u-g$ colors of the $z=0.1\pm0.027$ spectra against the LSST $g-r$ colors for the same spectra redshifted to $z=0.6\pm0.04$ and (right) the synthetic {\it GALEX}/SDSS $NUV-g$ colors of the $z=0.1\pm0.027$ spectra against the LSST $u-i$ colors for the same spectra redshifted to $z=0.6\pm0.04$. There is a systematic shift between the 1-to-1 line (dotted orange) and the best-fit linear regression (solid orange). This systematic must be corrected for so that the {\it GALEX}/SDSS--trained classifier can be applied to LSST data. Using the expected uncertainties in the photometric redshifts ($\sigma_z = 0.027$ at $z=0.1$ and $\sigma_z = 0.04$ at $z=0.6$), we estimate the likely uncertainties introduced in k-correcting the LSST colors to $z=0.1$. The rms about the best-fit lines shown here is 0.07 mag for $u-g$ and 0.09 mag for $NUV-g$, comparable to the measurement uncertainties. Thus, our machine learning model trained on {\it GALEX}/SDSS colors at $z\sim0.1$ can be applied to k-corrected LSST data to select likely TDE hosts at $z\sim0.6$.
}
\label{fig:synth_phot}
\end{figure*}

To quantify the impact of k-corrections on the performance of the Random Forest classifier, we estimate the likely uncertainties in k-corrected colors from the expected photometric redshift precision for LSST. For an accuracy of $\sigma_z/(1+z) \sim 0.025$ \citep{Salvato2018}, the redshift uncertainty will be $\sigma_z = 0.027$ at $z=0.1$ and $\sigma_z = 0.04$ at $z=0.6$. Next, we generate synthetic spectra of star-forming, quiescent, quiescent Balmer-strong, and post-starburst galaxies using FSPS (as in \S\ref{sec:models}). We generate two sets of redshifts, one with $z=0.1\pm0.027$ to model the training sample, and one with $z=0.6\pm0.04$ to model the performance of selecting on future galaxies from LSST, given the uncertainties in the photometric redshifts. In Figure \ref{fig:synth_phot}, we plot the synthetic SDSS $u-g$ colors of the $z=0.1\pm0.027$ spectra against the LSST $g-r$ colors for the same spectra redshifted to $z=0.6\pm0.04$. After fitting a linear relation, the residual uncertainty in the $u-g$ color is 0.07 mag, comparable to the typical measurement uncertainty of 0.07 mag for the SDSS $u-g$ colors. We also consider the $z=0.1\pm0.027$ SDSS $NUV-g$ colors vs. the $z=0.6\pm0.04$ LSST $g-r$ colors. The residual uncertainty after a linear fit is 0.09 mag, smaller than the typical measurement uncertainty of 0.12 mag for the {\it GALEX}/SDSS $NUV-g$ color.

If, to mimic the uncertainties introduced by k-correcting the LSST colors, we add the random errors of 0.07--0.09 mag above to the colors in the SDSS test sample, the purity and completeness of the Random Forest classification do not fall significantly. Thus, a machine learning model derived from the {\it GALEX}/SDSS training sample can be applied successfully to k-corrected LSST data to select likely TDE hosts at $z\sim0.6$.

\subsubsection{Are SDSS training redshifts appropriate for DES, Pan-STARRS, and photometric-only SDSS?}

In selecting galaxies from DES, Pan-STARRS, and SDSS photometry, we have assumed they are all at $z\sim0.1$, similar to the SDSS training/test sample, an assumption which we test here. The photometric datasets we consider from DES, Pan-STARRS, and SDSS are not as deep as LSST will be, and because we have required {\it WISE} data, the galaxies in the new catalogs discussed in \S\ref{sec:newcat} are likely within $z<0.5$. We thus do not attempt to use different observed SDSS bands by redshift as described for LSST in \S\ref{sec:models} and \S\ref{sec:lsst_error}. For the DES and PanSTARRS samples, we use the photometric transformations from \citet{Tie2017} and \citet{Tonry2012} to convert the data to photometric systems consistent with the SDSS training sample, so there will be no significant effect due to the differing filtersets.

We test the effect of the redshift uncertainties of the new DES, Pan-STARRS, and SDSS photometric samples on the purity and completeness of our selection, for different redshifts. We generate synthetic photometry for star-forming, early type, quiescent Balmer-strong, and post-starburst galaxy star formation histories at $z=0.1$, similar to the median of the SDSS sample \footnote{These synthetic training/test samples are only used for testing our methodology, and are not applied to generate a Random Forest classifier to be applied to real data.}. 

To simulate the performance on real galaxies, we add Gaussian noise to the synthetic photometry, with a $\sigma = 0.1$ mag. Using the synthetic training sample at $z=0.1$, and a scatter-added test sample at $z=0.3$, we find a purity of 66\% and a completeness of 79\%. Without redshifting (if both the synthetic training and test samples are at $z=0.1$), but with the same $\sigma = 0.1$ mag of scatter added to the synthetic test sample, the purity and completeness obtained are 68\% and 90\%, respectively. Without any scatter added to the test sample, using the training sample at $z=0.1$ and a test sample at $z=0.3$ results in a purity of 100\% and a completeness of 93\%. The effect of redshifting the synthetic training sample is thus much less than the effect of adding reasonable scatter. We expect then that our selection from the new photometric catalogs may have lower completeness than anticipated if the redshift cannot be estimated, but the purity (and thus the number of true quiescent Balmer-strong galaxies) will be as high as predicted from the SDSS training and test samples.

Not assuming likely redshifts for the new photometric samples will lead to the deterioration in completeness described above. This problem may cause the $\sim70$\% selection efficiency of the DES and Pan-STARRS samples(\S\ref{sec:newcat}).

\subsection{Absolute Magnitude-Dependent Effects}
\label{sec:magdep}

While we assume no evolution in galaxy properties between the SDSS training/test samples and the LSST, DES, PanSTARRS, and SDSS photometric surveys, we consider here whether differences in depth and volume lead to sampling different kinds of galaxies. We expect a larger number of intrinsically fainter galaxies due to the increased depth as well as a larger number of intrinsically brighter galaxies due to the increased volume of the new surveys. Such galaxies will be unlikely to mimic the colors of post-starburst or quiescent Balmer-strong galaxies, and should not significantly contaminate the samples selected via our photometric-only methodology.  As LSST photometric catalogs become available, our training and testing can be refined using data taken in the intervening years, such as with spectra from DESI.

\begin{figure*}
\includegraphics[width = 0.49\textwidth]{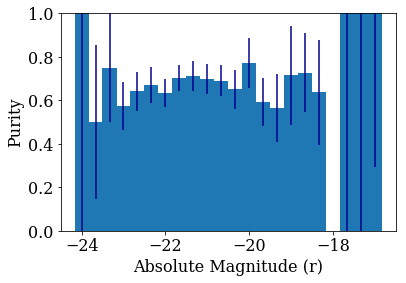}
\includegraphics[width = 0.49\textwidth]{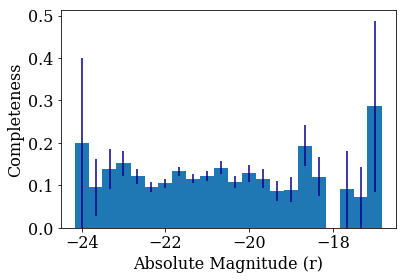}
\caption{Dependence of purity and completeness on the $r$-band absolute magnitude for the quiescent Balmer-strong galaxy selection from the SDSS test sample. There are no significant declines in purity or completeness with $M_r$, which implies that the selection will not suffer for intrinsically fainter or brighter objects just outside the range probed here (see \S\ref{sec:magdep}).}
\label{fig:mag_dependence}
\end{figure*}

We test the dependence of the purity and completeness on absolute $r$-band magnitude using the SDSS test/training samples (Figure \ref{fig:mag_dependence}). We do not observe a significant dependence of either purity or completeness on $M_r$, so the selection method is unlikely to suffer for objects just outside the range probed by the SDSS training/test samples.

We also test whether the $M_r$ distribution is different between the quiescent Balmer-strong galaxies used to train the Random Forest and those in the SDSS test sample that are successfully selected. No significant difference is observed, and a KS test shows we cannot reject the possibility that the $M_r$ distribution of each sample is the same. We do not expect the host galaxies of TDEs to differ in their $M_r$ distribution from the quiescent Balmer-strong galaxies, as we showed in \citet{French2017} that the volume corrected stellar mass function for the quiescent Balmer-strong galaxies selected from the SDSS was consistent with the eight TDE hosts considered there. However, that analysis was limited by small number statistics, and a difference may arise at black hole masses of $\gtrsim10^{8}$ M$_\sun$, and thus stellar masses of $\gtrsim10^{11}$ M$_\sun$, due to the tidal radius shrinking into the event horizon at typical black hole masses.

In summary, the absolute magnitudes of the SDSS training sample are appropriate for the LSST,  DES, PanSTARRS, and SDSS photometric surveys because the new intrinsically fainter or brighter galaxies are unlikely to contaminate the green valley and because the purity and completeness in the SDSS test sample do not depend on absolute magnitude.

\subsection{Improvements from Imminent Surveys}
\label{sec:futuredata}
In the years until LSST begins to detect transient events, more data will become available to improve the training samples used here. DESI \citep{desicollaboration2016} and BigBOSS \citep{schlegel2011} will provide deep spectra for over 10 million galaxies out to $z\sim0.2$, in addition to bright emission line galaxies, luminous red galaxies, and QSOs to $z\sim2$\footnote{\url{http://desi.lbl.gov}}\footnote{\url{http://bigboss.lbl.gov/}}. Wide optical surveys like DES, DECaLs, and BASS will provide corresponding photometry to $\sim$24th magnitude limits, several magnitudes deeper than SDSS. These surveys will increase the training samples of spectroscopically-identified post-starburst and quiescent Balmer-strong galaxies, as well as provide better photometric data to simulate what will be available with LSST. 

Higher redshift post-starburst galaxies can also be selected using photometric data, if enough bands are available; \citet{Maltby2016} use a PCA analysis \citep{Wild2014} to find $z\sim1$ post-starburst galaxies at 80\% purity with deep photometric data from the UKIDSS ultra-deep survey.

Complementing the galaxy surveys are upcoming transient surveys like the Zwicky Transient Facility (ZTF; \url{www.ptf.caltech.edu/ztf}), which will increase the number of TDEs detected and identified per year to $\sim30$ \citep{Hung2017}. These improved statistics will allow better characterization of their most distinguishing host galaxy properties, as well as of the rate evolution with redshift. The connection of TDEs with post-starburst and quiescent Balmer-strong star formation histories is likely to have some other underlying driver. For example, if the central stellar concentration is the determining factor \citep[see e.g.,][]{Stone2016, Graur2017, Law-Smith2017}, photometrically selecting these potential host galaxies may be a better option. However, the enhancement of the TDE rate appears to be higher in post-starburst and quiescent Balmer-strong hosts than in centrally-concentrated hosts \citep[to the resolution accessible with ground-based optical images;][]{Graur2017}, and contamination from other transients may make following up all transients in galaxies selected only by central concentration impractical. 

Our analysis here assumes that the TDE rate per galaxy (as well as the supernova rate) is constant with redshift. This is likely an oversimplification, which data from current and planned transient surveys can allow us to refine before LSST. Similarly, the rate of large AGN flares in quiescent Balmer-strong galaxies unassociated with TDEs, as well as the evolution of such events with redshift, will be constrained, as discussed in \S\ref{sec:othertransients}.

Many of the scenarios that explain the high TDE rate in post-starburst and quiescent Balmer-strong galaxies predict an even higher rate during the mid-late starburst phase, which might coincide with variable AGN activity, so such galaxies may host TDEs, which may be harder to detect given the increased brightnesses and dustiness of their cores, and the presence of AGN variability. Indeed, there are claimed TDE detections in both starbursting galaxies \citep{Tadhunter2017} and Seyferts \citep{Blanchard2017}. However, the supernova rates will also increase in these galaxies, and the details of the TDE and supernova rate comparison will affect their feasibility as good TDE identifiers. As more TDEs are found, the rates will become better understood as a function of their host galaxy properties; constantly updating the host galaxy information during LSST operations will be important in optimizing strategies for identifying TDEs.

\section{Conclusions}
Post-starburst and other quiescent Balmer-strong galaxies are the preferred hosts of Tidal Disruption Events. Thus, a transient detected in such a galaxy is interesting. Yet only a small fraction of potential TDE host galaxies has been identified spectroscopically. Here, we use machine learning to select such galaxies using photometric data alone. By applying a Random Forest classifier, we achieve a sample purity of 61\% and completeness of 21\% at $z\sim0.1$. At $z\sim0.5$, where we expect most TDEs in LSST to be detected, we achieve a sample purity of 53\% and completeness of 8\%. The subset of galaxies with the highest TDE rates, post-starbursts, are selected with a sample purity of 73\% and completeness of 12\% at $z\sim0.1$. The host galaxy observable features identified as most important to the Random Forest classification are the optical $u-g$ and $r-i$ colors and the {\it WISE} [4.6]--[12]$\mu$m color.

Assuming a purity of $\gtrsim$ 53\%, and the range of likely supernovae and TDE rates in quiescent Balmer-strong galaxies, we estimate that 13--99\% of transients observed in host galaxies photometrically selected by our method will be TDEs instead of supernovae. This may be an underestimate given that the ``false positives" classified as likely host galaxies are clustered near the edge of our selection criteria.

Following up all transients in LSST that occur in photometrically selected quiescent Balmer-strong galaxies will yield 119--248 TDEs per year given that 36--75\% of TDEs occur in such hosts, and assuming a photometric host selection completeness of 8\%. Identifying likely TDE hosts in advance allows for both rapid follow up of TDE candidates and a way to identify TDEs without restricting classification to their own expected photometric indicators. 

Using our technique, we present a new catalog of 67,484 new candidate quiescent Balmer-strong galaxies and 10,560 new candidate post-starburst galaxies selected from the SDSS, Pan-STARRS, DES and {\it WISE}. These new samples represent an increase of 3.5$\times$ and $6.3\times$, respectively, compared to previously known, spectroscopically identified samples. The number of new quiescent Balmer-strong galaxies is consistent with that expected from Pan-STARRS and DES, given the quiescent Balmer-strong galaxy rate in the spectroscopically-identified training and test samples---an independent check on the Random Forest classifier. Our methods and this catalog in particular are a significant new tool for studying galaxy evolution. Furthermore, given the high probability that a transient in one of these galaxies is a TDE, the approach and results here can be used to identify likely TDEs for follow up observations at the time of their discovery.

\acknowledgements
We thank the referee for useful comments which have improved this manuscript. We thank Daniel Eisenstein for his motivating question about post-starburst selection in photometric surveys. KDF is supported by Hubble Fellowship Grant HST-HF2-51391.001-A, provided by NASA through a grant from the Space Telescope Science Institute, which is operated by the Association of Universities for Research in Astronomy, Incorporated, under NASA contract NAS5-26555. AIZ acknowledges funding from NSF grant AST-0908280 and NASA grant ADP-NNX10AE88G.

Based on observations made with the NASA Galaxy Evolution Explorer. GALEX is operated for NASA by the California Institute of Technology under NASA contract NAS5-98034.  

This publication makes use of data products from the Wide-field Infrared Survey Explorer\citep{2010AJ....140.1868W}, which is a joint project of the University of California, Los Angeles, and the Jet Propulsion Laboratory/California Institute of Technology, funded by the National Aeronautics and Space Administration. 
 
 The Pan-STARRS1 Surveys (PS1) have been made possible through contributions of the Institute for Astronomy, the University of Hawaii, the Pan-STARRS Project Office, the Max-Planck Society and its participating institutes, the Max Planck Institute for Astronomy, Heidelberg and the Max Planck Institute for Extraterrestrial Physics, Garching, The Johns Hopkins University, Durham University, the University of Edinburgh, Queen's University Belfast, the Harvard-Smithsonian Center for Astrophysics, the Las Cumbres Observatory Global Telescope Network Incorporated, the National Central University of Taiwan, the Space Telescope Science Institute, the National Aeronautics and Space Administration under Grant No. NNX08AR22G issued through the Planetary Science Division of the NASA Science Mission Directorate, the National Science Foundation under Grant No. AST-1238877, the University of Maryland, and Eotvos Lorand University (ELTE). 
 
 Funding for SDSS-III has been provided by the Alfred P. Sloan Foundation, the Participating Institutions, the National Science Foundation, and the U.S. Department of Energy Office of Science. The SDSS-III web site is http://www.sdss3.org/. SDSS-III is managed by the Astrophysical Research Consortium for the Participating Institutions of the SDSS-III Collaboration including the University of Arizona, the Brazilian Participation Group, Brookhaven National Laboratory, University of Cambridge, Carnegie Mellon University, University of Florida, the French Participation Group, the German Participation Group, Harvard University, the Instituto de Astrofisica de Canarias, the Michigan State/Notre Dame/JINA Participation Group, Johns Hopkins University, Lawrence Berkeley National Laboratory, Max Planck Institute for Astrophysics, Max Planck Institute for Extraterrestrial Physics, New Mexico State University, New York University, Ohio State University, Pennsylvania State University, University of Portsmouth, Princeton University, the Spanish Participation Group, University of Tokyo, University of Utah, Vanderbilt University, University of Virginia, University of Washington, and Yale University.  
 
 This project used public archival data from the Dark Energy Survey (DES). Funding for the DES Projects has been provided by the DOE and NSF (USA), MISE (Spain), STFC (UK), HEFCE (UK). NCSA (UIUC), KICP (U. Chicago), CCAPP (Ohio State), MIFPA (Texas A\&M), CNPQ, FAPERJ, FINEP (Brazil), MINECO (Spain), DFG (Germany) and the collaborating institutions in the Dark Energy Survey, which are Argonne Lab, UC Santa Cruz, University of Cambridge, CIEMAT-Madrid, University of Chicago, University College London, DES-Brazil Consortium, University of Edinburgh, ETH Zurich, Fermilab, University of Illinois, ICE (IEEC-CSIC), IFAE Barcelona, Lawrence Berkeley Lab, LMU Munchen and the associated Excellence Cluster Universe, University of Michigan, NOAO, University of Nottingham, Ohio State University, OzDES Membership Consortium, University of Pennsylvania, University of Portsmouth, SLAC National Lab, Stanford University, University of Sussex, and Texas A\&M University. Based in part on observations at Cerro Tololo Inter-American Observatory, National Optical Astronomy Observatory, which is operated by the Association of Universities for Research in Astronomy (AURA) under a cooperative agreement with the National Science Foundation.

This work made use of the IPython package \citep{PER-GRA:2007}. This research made use of Scikit-learn \citep{scikit-learn}. This research made use of SciPy \citep{jones_scipy_2001}. This research made use of TOPCAT, an interactive graphical viewer and editor for tabular data \citep{2005ASPC..347...29T}.  This research made use of Astropy, a community-developed core Python package for Astronomy \citep{2013A&A...558A..33A}. This research made use of NumPy \citep{van2011numpy}. This research made use of database access and other data services provided by the NOAO Data Lab.

\bibliographystyle{apj}
\bibliography{earefs.bib}

\end{document}